\documentclass[aps,prl,twocolumn,showpacs,superscriptaddress,groupedaddress]{revtex4}

\usepackage{graphicx}
\usepackage{dcolumn}
\usepackage{bm}
\usepackage{epsfig}
\usepackage{epstopdf}
\usepackage{amssymb}   
\usepackage{gensymb}
\usepackage{amsmath}
\usepackage{float}
\usepackage{color,soul}

\begin{document}

\widetext

\title{Enhanced Kadowaki-Woods Ratio and Weak-Coupling Superconductivity in Noncentrosymmetric YPt$_2$Si$_2$ Single Crystals}

\author{Gustavo Gomes Vasques}
\affiliation{Centro Brasileiro de Pesquisas F{\'i}sica (CBPF), Rio de Janeiro, RJ, 22290-180, Brazil.}
\affiliation{Centro de Ci{\^e}ncias Naturais e Humanas (CCNH), Universidade Federal do ABC (UFABC), Santo Andr{\'e}, SP, Brazil.}
\author{Shyam Sundar}
\affiliation{Instituto de F{\'i}sica, Universidade Federal do Rio de Janeiro, 21941-972 Rio de Janeiro, RJ, Brazil.}
\author{Deisy Aristiz{\'a}bal-Giraldo}
\affiliation{Instituto de F{\'i}sica, Universidad de Antioquia UdeA, Calle 70 No 52-21, Medell{\'i}n, Colombia.}
\affiliation{Departamento de F{\'i}sica y Astronom{\'i}a, Facultad de Ciencias Exactas, Universidad Andr{\'e}s Bello, Sazi{\'e} 2212, Santiago, Chile.}
\author{Juan F. Castello-Arango}
\affiliation{Instituto de F{\'i}sica, Universidad de Antioquia UdeA, Calle 70 No 52-21, Medell{\'i}n, Colombia.}
\author{Rafael S{\'a} de Freitas}
\affiliation{Instituto de F{\'i}sica, Universidade de S{\~a}o Paulo, 05508-090 S{\~a}o Paulo, SP, Brazil.}
\author{Adriano Reinaldo Vi\c{c}oto Benvenho}%
\affiliation{Centro de Ci{\^e}ncias Naturais e Humanas (CCNH), Universidade Federal do ABC (UFABC), Santo Andr{\'e}, SP, Brazil.}
\author{Takahiro Onimaru}
\affiliation{Department of Quantum Matter, Graduate School of Advanced Science and Engineering, Hiroshima University, Higashi-Hiroshima, Japan.}
\author{Jorge M. Osorio-Guill{\'e}n}
\affiliation{Instituto de F{\'i}sica, Universidad de Antioquia UdeA, Calle 70 No 52-21, Medell{\'i}n, Colombia.}
\author{Marcos A. Avila}
\affiliation{Centro de Ci{\^e}ncias Naturais e Humanas (CCNH), Universidade Federal do ABC (UFABC), Santo Andr{\'e}, SP, Brazil.}

\date{\today}

\begin{abstract}

Superconductivity in noncentrosymmetric RPt$_2$Si$_2$ (R = rare earth) compounds exhibit a rich playground to explore the competition between different ground states, such as unconventional superconductivity, antiferromagnetism and charge density wave. Here, we report the successful single crystal synthesis of noncentrosymmetric YPt$_2$Si$_2$ superconductor, with a transition temperature $T_c = 1.67$~K, via Sn flux method. The high quality of the prepared single crystals was confirmed using powder and Laue X-ray diffraction (XRD) measurements. The superconducting and normal-state properties are investigated using electrical transport and heat capacity measurements down to 0.5~K. In the normal state, unlike LaPt$_2$Si$_2$, no charge density wave (CDW) transition is observed in YPt$_2$Si$_2$, as evidenced by electrical transport and specific heat measurements. A relatively large Kadowaki-Woods ratio (KWR), $A/\gamma^2 = 5.17 \times 10^{-5}$ $\mu\ohm$-cm~(mol-K/mJ)$^2$, and a linear temperature variation of the electrical resistivity $\rho(T)$ in an extended temperature range of 50-300~K suggest an unconventional normal-state in YPt$_2$Si$_2$. The estimated superconducting parameters indicate that YPt$_2$Si$_2$ is a type-II superconductor with weak electron-phonon coupling. The temperature dependence of specific heat in the superconducting state can be explained reasonably well using an isotropic two-gap model. A positive curvature near $T_c$ in the temperature variation of upper critical field $H_{c2}^{\parallel c}(T)$ also supports the two-gap superconductivity. First-principles DFT calculations suggest a BCS-like superconducting state driven primarily by $d$-electron contributions. The calculated electron-phonon coupling constant($\lambda_{ep}$) identifies the material as a weak-coupling superconductor, with the McMillan-Allen-Dynes formula yielding a $T_c$ of 1.8 K. Additionally, we provide a comparative analysis of the superconducting and normal-state properties of YPt$_2$Si$_2$ and compositionally similar LaPt$_2$Si$_2$.

\end{abstract}

\maketitle


\section{\label{sec:level1}Introduction}
Superconductivity in materials with missing inversion symmetry has been continuously of interest since the discovery of unconventional superconductivity in noncentosymmetric CePt$_3$Si heavy-fermion compound \cite{CePt3Si_2004}. The lack of inversion symmetry in noncentrosymmetric superconductors (NCS) leads to an antisymmetric spin-orbit coupling (ASOC), which may give rise to a mixture of spin-singlet and spin-triplet pairing states of the superconducting wave function. Due to this, a multitude of unconventional properties are possible in NCSs, such as nodes in the superconducting gap, topological superconductivity, magneto-electrical effects, and time-reversal symmetry breaking (TRSB) superconducting state \cite{Naskar2021, Smidman_2017, bauer2012non}. In the quest for a comprehensive understanding, a wide variety of NCSs consisting of strong and weak electronic correlations have been studied over the years \cite{Naskar2021, Smidman_2017, bauer2012non}. However, NCSs with weak electronic correlations are of particular interest to investigate the unique role of ASOC in these materials. Several NCSs with weak electronic correlations have shown unconventional superconducting behaviour, such as the nodal superconducting gap structure in ThCoC$_2$ and LaPtSi \cite{ThCoC2_2019, LaPtSi_Shang2022}, and two-gap TRSB superconductivity in LaNiC$_2$ \cite{Sundar2021}. Moreover, the role of different crystal structures on the nature of superconductivity in NCS is yet to be established. 

\begin{figure}[]
\includegraphics[scale=0.45]{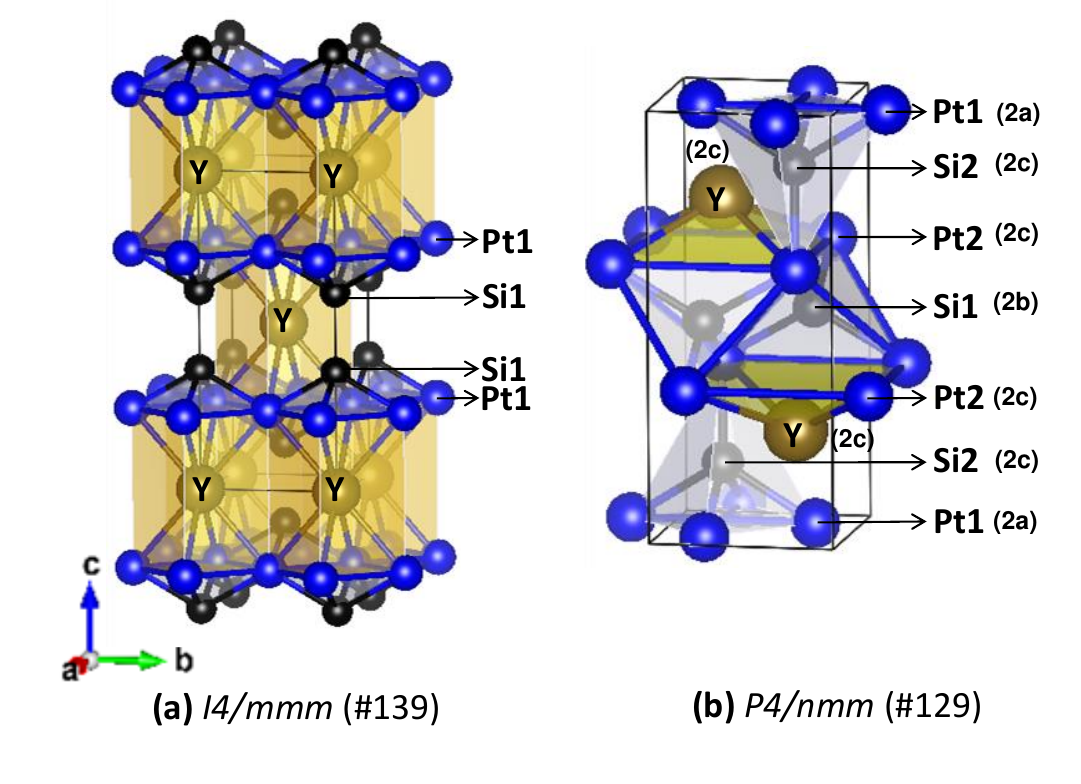}
\caption{\label{fig:structure} Two dominant structure types for RT$_2$X$_2$ compounds: (a) the centrosymmetric ThCr$_2$Si$_2$-type structure (\textit{I}4\textit{/mmm}) and (b) the noncentrosymmetric CaBe$_2$Si$_2$-type structure (\textit{P}4\textit{/nmm}).  Atomic representations: Y (dark gold), Si (black), and Pt (blue).}
\end{figure}

In general, ternary RT$_2$X$_2$ (R = rare earth, T = transition metals, and X = Si, Ge) compounds crystallize in the centrosymmetric ThCr$_2$Si$_2$-type structure (space group \textit{I}4\textit{/mmm} Fig.~\ref{fig:structure}(a)) \cite{SZYTULA1989133_compendio}, with the exceptions of T = Pt, which forms in the noncentrosymmetric CaBe$_2$Si$_2$-type structure (\textit{P}4\textit{/nmm} Fig.~\ref{fig:structure}(b)). This suggests that the lack of inversion symmetry in these compounds favors the superconducting state \cite{SHELTON1984797}. This argument is further supported by the observation of superconductivity in noncentrosymmetric polymorphs of YIr$_2$Si$_2$ and LaIr$_2$Si$_2$ compounds at $T_c=2.52$~K, and 1.24~K respectively \cite{YIr2Si2_2012}, for which no superconductivity was observed in their centrosymmetric counterparts. 

Noncentrosymmetric compounds in the RPt$_2$Si$_2$ (R = rare-earth) family, containing light rare-earth elements such as R = La, Pr, and Nd, have been of great interest due to the existence of multiple correlated phenomena, such as charge density wave, magnetism and superconductivity \cite{Nagano2013}. Among them, the LaPt$_2$Si$_2$ compound has been studied in detail to investigate the complex interplay of the charge density wave and superconductivity with a $T_c \approx 1.8$~K and $T_{CDW} \approx 85$~K \cite{Gupta_2017_La, Falkowski_2019_La, La_muon_2018, La_pressure_2020, La_RIXS_2022, La_syncrotron_NOCERINO2023100621}. The hydrostatic pressure tuning of the CDW and superconducting state shows the competing nature of the two electronic states, where the maximum $T_c$ is observed at a pressure where the CDW order is fully suppressed \cite{La_pressure_2020}. Superconducting-gap structure investigated using several techniques, including muon spin rotation ($\mu$SR) \cite{La_muon_2018}, tunnel diode oscillator \cite{Nie2021}, specific heat \cite{La_pressure_2020} and soft point contact spectroscopy \cite{Nie2021}, suggests either single or two-gap nodeless superconductivity. Two sister compounds of LaPt$_2$Si$_2$, namely SrPt$_2$Si$_2$ \cite{SrPt2As2_2010}, and BaPt$_2$Si$_2$ \cite{Guo2016} also display the superconducting and CDW transitions. Therefore, it is natural to search for other members of this family to further investigate and obtain a comprehensive picture of the interplay between CDW and superconductivity. The YPt$_2$Si$_2$ compound has been synthesized previously in the polycrystalline form using the arc-melting technique which shows a superconducting transition at $T_c \approx ~ 1.6$~K, but no CDW transition \cite{Pikul_2017}. It is generally known that disorder may suppress the signature of a CDW transition, therefore an investigation of physical properties in single crystal of YPt$_2$Si$_2$ is important.

Here, we report the successful synthesis of YPt$_2$Si$_2$ single crystal using Sn flux. The high quality of the crystals is confirmed through powder X-ray diffraction (XRD), as well as Laue XRD measurements. Both XRD data sets confirmed the CaBe$_2$Si$_2$-type structure of the YPt$_2$Si$_2$ crystals. Superconducting and normal state properties were studied using the temperature dependence of electrical resistivity and specific heat measurements down to 0.5 K. Experimental results were complemented with detailed theoretical calculations, based on density functional theory (DFT), providing the electronic band structure, Fermi surface, and phonon density of states. 

\section{Methodology}

\subsection{Experimental}

Single crystals of YPt$_2$Si$_2$ were grown by the Sn-flux method. A mixture of the starting materials, Y (99.995$~\%$), Pt (99.95+$~\%$), Si (99.999$~\%$), and Sn (99.999$~\%$), in a molar ratio of 1:2.5:1.5:95, was placed in a quartz ampoule and sealed under vacuum. The ampoule was initially heated  to 500~$\degree$C, and held at this temperature for 1 hour. It was then heated to 1180~$\degree$C over a period of 3 hours and maintained at that temperature for 24 hours to ensure a homogeneous melt of the reagents. Subsequently, the ampoule was cooled down to 680~$\degree$C at a constant rate of 2~$\degree$C/h, after which it was removed from the furnace and immediately centrifuged to remove the excess Sn flux. Residual Sn flux on the crystals was removed using hydrochloric acid (HCl). The typical dimensions of the resulting single crystals are approximately 0.9$~\times~$0.4$~\times~$0.08$~$mm$^3$. 

Structural characterization of the single crystals were performed with the help of powder X-ray diffraction (XRD) (diffractometer model: STOE STADI-P) using Cu-K$\alpha$ ($\lambda = 1.5406$~{\AA}) radiation. The high crystalline quality of the single crystals were also attested by the XRD data from a Laue diffractometer (Photonic Science). The stoichiometric composition of the crystals was confirmed using dispersive X-ray spectroscopy (EDX) coupled with a scanning electron microscopy (SEM) system (JEOL, JSM-6010LA). The average molar stoichiometry obtained by EDX measurements is found to be Y~=~0.89(4), Pt~=~2.03(9), Si~=~2.0(1). Electrical resistivity and specific heat measurements were performed down to 1.8~K and extended down to 0.4~K with a $^3$He insert, using a Quantum Design (QD) Physical Property Measurements System (PPMS). AC magnetic susceptibility measurements were performed using a home-built susceptometer operating at an excitation frequency of 155 Hz with a field amplitude of 1.8 Oe, employing the mutual inductance bridge technique in a pumped $^4$He cryostat.

\subsection{Theoretical}

Spin-polarized first-principles calculations were performed within the framework of DFT. Spin-orbit coupling (SOC) was incorporated using the projector augmented-wave (PAW) method \cite{blochl1994projector} as implemented in the Vienna Ab initio Simulation Package (VASP) \cite{kresse1999ultrasoft,hobbs2000fully}. The exchange-correlation potential was treated using the Perdew--Burke--Ernzerhof functional revised for solids (PBEsol) \cite{perdew2008restoring}. The electronic wave-functions were expanded in a plane-wave basis set with a kinetic energy cutoff of 400~eV, including 160 conduction bands. Electronic convergence was reached when the total energy change between successive iterations was less than $10^{-8}$~eV. Structural relaxations utilized a $\Gamma$-centered $12\times12\times6$ $\mathbf{k}$-point mesh, while a denser $16\times16\times8$ grid was employed to calculate the electronic band dispersion, the total and site-projected density of states (DOS), and the Fermi surface (FS). Phonon properties and electron--phonon coupling (EPC) calculations were performed without SOC using Density Functional Perturbation Theory (DFPT) \cite{RevModPhys.73.515} as implemented in the Quantum ESPRESSO (QE) suite \cite{Giannozzi2017_QE,Giannozzi2009_QE}. For these calculations, PAW datasets from the PSLibrary (v.1.0.0) \cite{DalCorso2014_PSPTable} were used, with plane-wave and charge-density cutoffs set to 90~Ry and 900~Ry, respectively. Electronic occupations were treated using a Gaussian smearing of 0.02~Ry. The phonon Brillouin zone (BZ) was sampled with a $6\times6\times3$ $\mathbf{q}$--grid, and the self-consistent DFPT equations were solved until the squared norm of the first-order potential residual was below $10^{-12}$. EPC matrix elements were obtained by interpolating the dynamical matrices from a coarse $18\times18\times12$ $\mathbf{k}$--mesh to a finer $36\times36\times24$ grid \cite{Wierzbowska2006_NbTc_arXiv}. The phonon DOS was calculated using the tetrahedron method \cite{Blochl1994_Tetrahedron}. To evaluate the Eliashberg spectral function $\alpha^2F(\omega)$ and $\lambda_{ep}$, which involve double-delta integrations over the FS, a Gaussian broadening technique was applied \cite{Wierzbowska2006_NbTc_arXiv}. Finally, the superconducting transition temperature ($T_c$) was estimated via the McMillan--Allen--Dynes formula, assuming a semi-empirical Coulomb pseudopotential of $\mu^{*}=0.13$.

\section{Experimental Results}

\subsection{Crystal Structure}

YPt$_2$Si$_2$ crystallizes in a tetragonal noncentrosymmetric structure with space group $P4/nmm$, as shown in Fig.~\ref{fig:structure}(b). It has 10 atoms per primitive cell with a single Wyckoff position for the Y atom and two non-equivalent sites for the Pt and Si atoms. There are Pt1 and Si1 at a $2a$ $(-4m2)$ and $2b$ $(-4m2)$ sites, respectively, while Y, Pt2 and Si2 are localized at a $2c$ $(4mmm)$ sites. Table \ref{tab:XRD_parameters} lists the Wyckoff positions for each atom. The structure can be described in terms of interconnected triangular and square pyramids. Two triangular pyramids, centered within the unit cell, are formed by Pt2 atoms forming isosceles triangles, which enclose Si1 atoms. These triangular pyramids connect to two square pyramids via shared Pt2 atoms, which form the bases of the square pyramids with Y at their apices. The remaining two square pyramids, located at the corners of the unit cell, consist of Pt1 atoms forming the bases and Pt2 at the apex, with Si2 encapsulated within. Interatomic distances within these polyhedra are provided in Table~II.

\begin{figure}[h]
\centering
\includegraphics[scale=0.3]{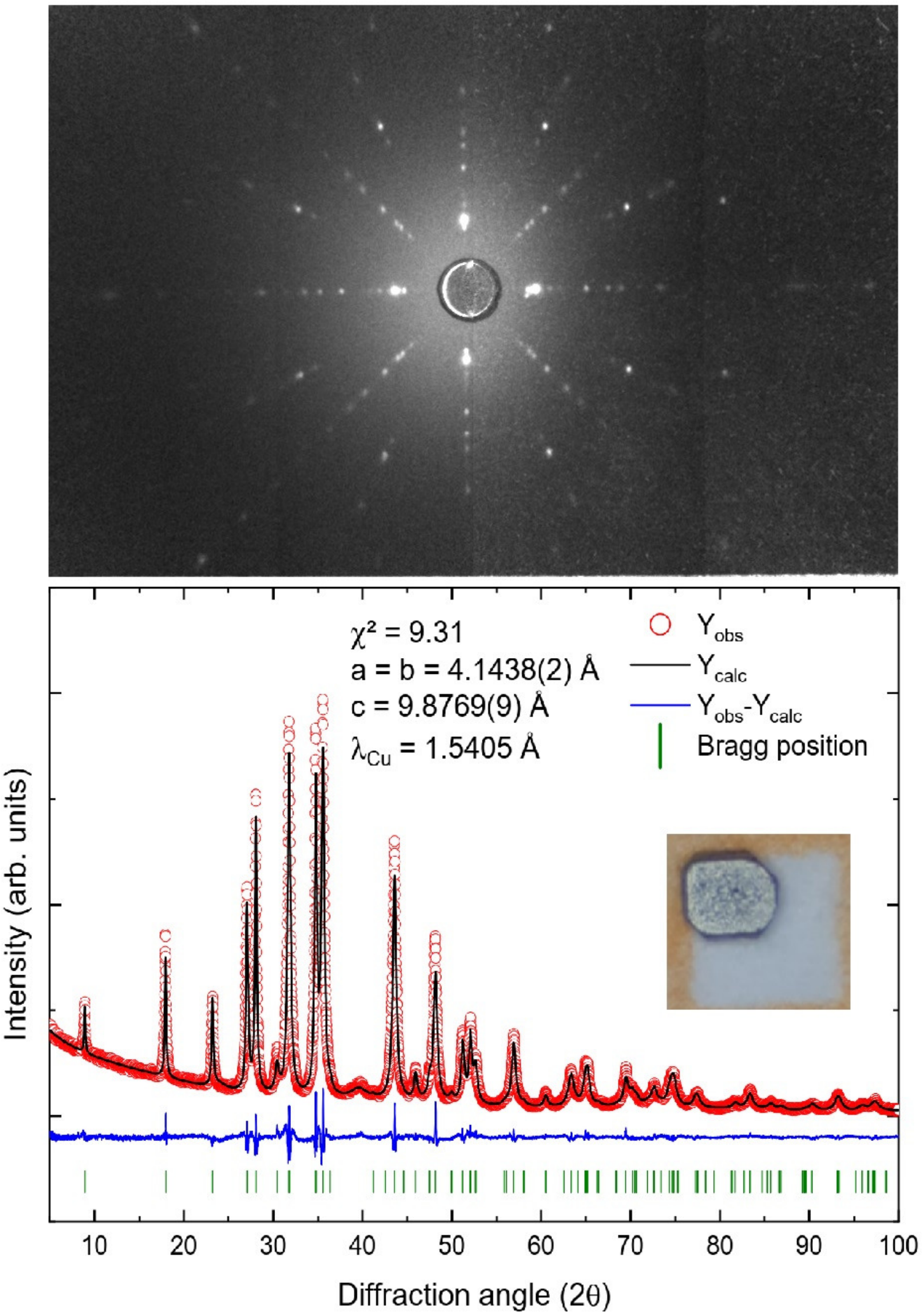}
\caption{\label{fig:DRX}Upper) Laue photograph along the (001) plane, and bottom) powder XRD pattern at room temperature. The solid black lines represent the Rietveld refinement fit, the solid blue line represents the difference between the observed and calculated profile and the vertical green lines shows the Bragg positions.}
\end{figure}

The upper panel of Fig. \ref{fig:DRX} shows the Laue backscattering pattern of YPt$_2$Si$_2$ crystal along (001) direction. The well defined diffraction spots with the tetragonal symmetry confirm the high crystalline quality of the crystals. Moreover, the Lauegram revealed that the plate-like crystals grow along the ab-axis, whereas the thickness is aligned with the c-axis. The powder XRD pattern of a crushed YPt$_2$Si$_2$ crystal is shown in the lower panel of Fig. \ref{fig:DRX}. The powder XRD data is successfully fitted using the Rietveld refinement method considering CaBe$_2$Si$_2$-like structure (space group: \textit{P}4\textit{/nmm}). A small hump observed at $\sim 40\degree$ is due to the glue used to hold the sample in the sample holder for the Debye-Scherrer geometry. The lattice parameters obtained from the Rietveld refinement are $a = b = 4.1438$~{\AA}, and $c = 9.8769$~{\AA}. These values are in good agreement with the literature on polycrystalline YPt$_2$Si$_2$ \cite{Pikul_2017, Nagano2013, HIEBL1985}. In comparison, the volume of the unit cell of YPt$_2$Si$_2$ is $\approx$ 5~$\%$ smaller than LaPt$_2$Si$_2$ \cite{Falkowski_2019_La}.

\begin{table}[]
\begin{ruledtabular}
\caption{\label{tab:XRD_parameters}
Lattice parameters, unit cell volumes and atomic coordinates of the single crystalline YPt$_2$Si$_2$ at 300~K.}    

\begin{tabular}{c|c|ccc}
\hline
Lattice parameters & Atom site & \multicolumn{3}{c}{Position}                                 \\ \hline
\textit{P}4\textit{/nmm} (\#129)     &           & \multicolumn{1}{c|}{x}    & \multicolumn{1}{c|}{y}    & z      \\ \hline
$a = 4.1438(2)$ \AA     & Y (2\textit{c})    & \multicolumn{1}{c|}{0.25} & \multicolumn{1}{c|}{$0.25$} & $0.7488(3)$\\ \hline
$c = 9.8769(9)$ \AA     & Pt1 (2\textit{a})  & \multicolumn{1}{c|}{$0.75$} & \multicolumn{1}{c|}{$0.25$} & 0 \\ \hline
$V = 169.60(4)$ \AA$^3$      & Pt2 (2\textit{c})  & \multicolumn{1}{c|}{0.25} & \multicolumn{1}{c|}{0.25} & 0.3752(9)      \\ \hline
                   & Si1 (2\textit{b})  & \multicolumn{1}{c|}{0.75} & \multicolumn{1}{c|}{0.25} & 0.5 \\ \hline
$\chi ^2 = 9.31$        & Si2 (2\textit{c})  & \multicolumn{1}{c|}{0.25} & \multicolumn{1}{c|}{0.25} & 0.1204(4)     \\ \hline
\end{tabular}
\end{ruledtabular}
\end{table}

\begin{table}[h!]
\caption{Polyhedral information (in \AA) for YPt$_2$Si$_2$}
    \centering
\begin{tabular}{lc}
\hline
\textbf{Isoceles triangular pyramids} & \\ 
triangle lengths: & 3.777 and 4.150 \\
\\
\textbf{square pyramids 1} &\\
base lengths: & 4.162 \\
apex atom Y-Pt2 distance: & 3.173 \\
\\
\textbf{square pyramids 2} & \\
base lengths: & 2.934\\
apex atom Pt1-Pt2 distance: & 4.282 \\
\hline
\end{tabular}
\label{}
\end{table}

\begin{figure*}[t]
\centering
\includegraphics[scale=0.35]{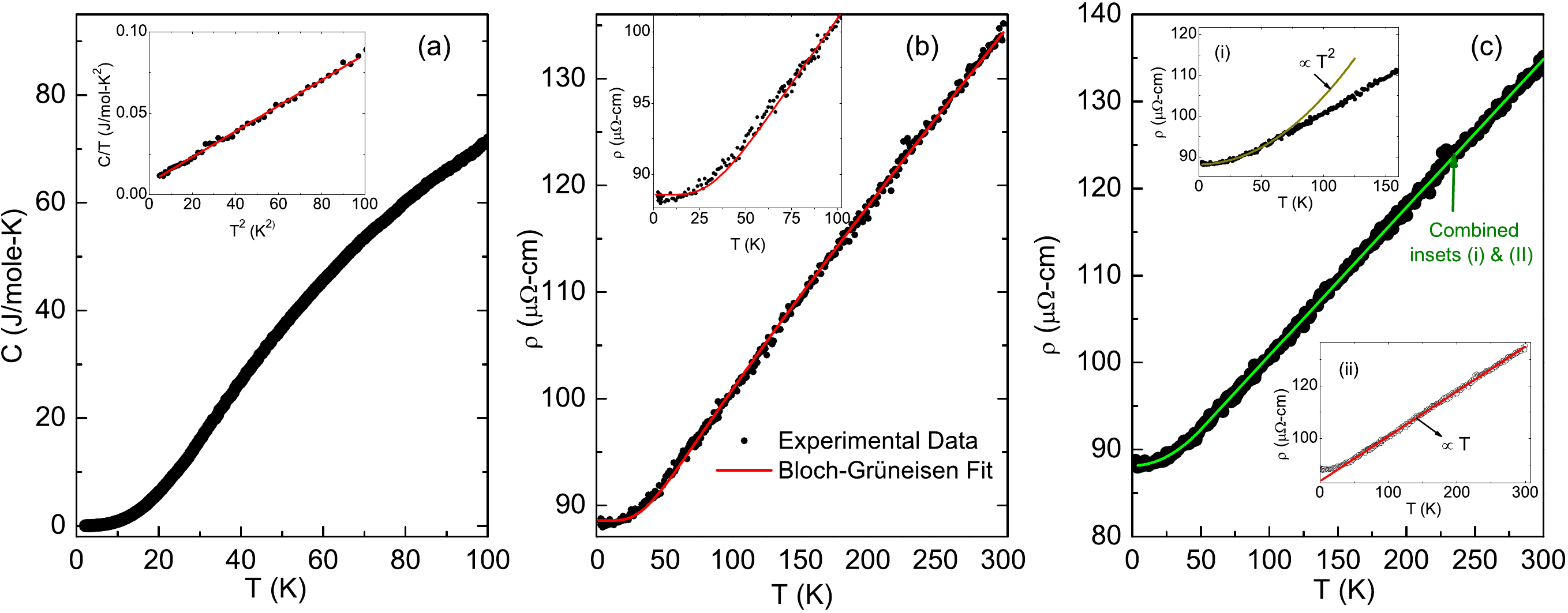}\\
\caption{\label{fig:normal} (a) Temperature dependence of specific heat, $C(T)$, at zero applied magnetic field in the range 2-100 K. Inset shows a linear fit of $C_p/T$ vs. $T^2$ in the $T$ range of 2-10 K (see details in the text). (b) Temperature dependence of electrical resistivity, $\rho(T)$, in the $T$ range of 2-300 K. Solid line represents the fit of the experimental data using the Bloch-Gr{\"u}neisen (BG) model. Inset shows that the BG model can not explain the data satisfactorily below $\approx$ 75 K (see main text for details). (c) $\rho(T)$ in the $T$ range of 2-300 K. The solid line represents the combined fit using a linear fit at high temperature (50 - 300 K) and a $T^2$ dependence in the range 2 - 50 K. Inset (i) and (ii) show the $T^2$ and linear temperature dependence of $\rho(T)$ in the respective temperature ranges.}
\end{figure*}

\subsection{Normal state}

The temperature dependence of the specific heat, in the temperature range 2-100~K, is shown in the main panel of Fig. 3a. Contrary to LaPt$_2$Si$_2$, no CDW transition is observed in the specific heat below 100~K for YPt$_2$Si$_2$. In the low-temperature range from 2-10~K, the specific heat data is fitted using $C(T)=\gamma T+\beta T^3$, where $\gamma$ is the coefficient of electronic specific heat and $\beta$ allows extraction of the value of the characteristic Debye temperature, $\theta_D$. The first and the second term in the expression represent the electron and phonon contributions of specific heat. A straight line fit to the $C/T$ vs. $T^2$ plot (in the inset of Fig. 3a) provides $\gamma=7.3(5)~$mJ/mol$\cdot$K$^2$ and $\beta=0.8(1)~$mJ/mol$\cdot$K$^4$. The Debye temperature, $\theta_D = 228(4)$~K, agrees well with the one obtained for polycrystalline YPt$_2$Si$_2$ in the  literature \cite{Pikul_2017}.
 \begin{equation}
    \theta_D = \left(\frac{12\pi^4 n R}{5 \beta}\right)^\frac{1}{3}
\end{equation}

The temperature dependence of electrical resistivity, $\rho$($T$), at zero applied magnetic field, is shown in Fig. 3(b). Although the absolute value of the residual resistivity in the single crystal has been reduced by half, compared to polycrystalline samples \cite{Pikul_2017}, it still shows a small residual resistance ratio (RRR) $\approx$~1.52~\cite{Pikul_2017}. Such small values of RRR have been observed in other RPt$_2$Si$_2$ compounds, similar to iron-pnictide superconductors, suggesting a natural tendency of crystallographic disorder in this series \cite{LuT2Si2_2018, Rafael2022}. The experimentally determined residual resistivity and the Sommerfeld coefficient have been used to estimate the mean free path $l = v_F \tau$, where $v_F = \hbar k_F/m^*$ is the Fermi velocity, and $\tau = m^*/ne^2\rho_0$ is the scattering time constant from the Drude model. By assuming a spherical Fermi surface ($k_F = \sqrt[3]{3n\pi^2})$ and the electron density $n$ being calculated using the contribution of the three electrons from $Y^{3+}$, as there are two formula units per unit cell of the compound ($Z = 2)$ the electron density is $n = 6/V_{cell} = 3.53 \times 10^{28}~$m$^{-{3}}$ and $m^*= \hbar^2 k_F^2 \gamma / \pi^2 n k_B^2 = 2.71m_0$, where $m_0$ is the free-electron mass and $\gamma$ is in volume units. This leads to a Fermi velocity $v_F = 4.33 \times 10^5~$m/s and mean free path $l = 1.34$~nm.

In Fig.~3(b), the $\rho$($T$) in the temperature range 2-300~K is fitted using the Bloch-Gr\"{u}neisen (BG) model: 
\begin{equation}
		\rho = \rho_0+B\left(\frac{T}{\theta_{BG}}\right)^5\int_{0}^{\theta_{BG}/T} \frac{x^5dx}{(e^x - 1) (1- e^{-x})},
\end{equation}
where $\rho_0$ is the residual resistivity due to the static disorder, the second term stands for the electron-phonon scattering, where $B$ is related to the electron-phonon coupling, and $\theta_{BG}$ the Debye temperature obtained from resistivity. The parameters obtained from the least-square fitting are $\rho_0 = 88.58~\mu \Omega$-cm, $B = 132.98~\mu \Omega$-cm, and $\theta_{BG} = 212$~K \cite{Hwang_2019, ziman2001electrons}. Although the BG model explains the $\rho$($T$) data well above 75~K, it fails to capture the low temperature behaviour (see inset of Fig. 3(b)). This suggests that the electron-phonon scattering ($T^5$-law) does not play significant role at low temperature in the electron transport mechanism in YPt$_2$Si$_2$. However, the $\rho$($T$) at low temperature (inset (i) of Fig. 3(c)), in the range of 2-50~K, can be well explained using, $\rho_0 + A T^2$, where $\rho_0$ is the residual resistivity, and $A$ is the coefficient of $T^2$ term responsible for electron-electron scattering mechanism. The fitted values of parameters $\rho_0$ and $A$ are 88.16(5)~$\mu \Omega$-cm, and 1.66(4)~$\times$ 10$^{-3}$~$\mu \Omega$-cm/K$^2$ respectively. Moreover, above 50~K, the $\rho$($T$) data follows a linear temperature dependence up to 300 K, with a slope of 0.1701(3)~$\mu \Omega$-cm/K, shown in the inset (ii) of Fig. 3(c). A resultant fitted curve combining the linear and quadratic temperature variation of $\rho$($T$) is shown in the main panel of Fig. 3(c). Such a linear variation of electrical resistivity in a broad temperature range is not common, and is widely classified as ``strange metal'' phase. Such strange metal phases generally emerge in heavy-fermion, quantum critical and low-carrier density quantum materials \cite{Bruin_2013}. However, none of the above phenomena has been observed in YPt$_2$Si$_2$. In general, according to the Boltzmann transport theory, a linear variation of $\rho$($T$) can be observed above $T>\theta_D$ due to low energy phonon scattering, as reported for TlBi$_2$ \cite{Yang_2022} and for SrPt$_3$P \cite{Takayama_2012}. In the case of YPt$_2$Si$_2$, the $\theta_D$ estimated from the specific heat and electrical resistivity measurements are 228~K and 212~K respectively. Therefore, electron-phonon scattering alone cannot explain the linear variation of $\rho$($T$) down to $\approx$~50~K in YPt$_2$Si$_2$. This suggests the possibility of electron-electron interaction mechanism in YPt$_2$Si$_2$ similar to what is observed in heavy heavy-fermion compounds \cite{Bruin_2013}, but the exact origin of the $T$-linear resistivity in YPt$_2$Si$_2$ is yet to be established.   

To gain insight into the electron correlations in YPt$_2$Si$_2$, we compared the temperature dependence of the electrical resistivity and the heat capacity. From this analysis, we obtained the Kadowaki-Woods ratio (KWR), $A/\gamma^2 \approx 5.17 \times 10^{-5} \mu\Omega$-cm~(mol-K/mJ)$^2$. In general, the KWR probes the strength of electron-electron scattering and the renormalization of the effective electron mass. For many heavy-fermion materials, the KWR is approximately $1.0 \times 10^{-5}\mu\Omega$-cm~(mol-K/mJ)$^2$, as shown in Fig. 4. Since YPt$_2$Si$_2$ exhibits a relatively small $\gamma$, the observation of a KWR larger than that of typical heavy-fermion materials is rather surprising. Similarly enhanced KWR values have also been reported in other superconducting systems that display low $\gamma$, such as Nb$_{0.18}$Re$_{0.82}$, NaAlSi, and TlBi$_2$ \cite{Sundar_2019, Yamada_2021, Yang_2022}. A theoretical framework proposed independently by Yu and Andersson \cite{Yu_1984} and by Matsuura and Miyake \cite{Matsuura_1986} suggests that a strong dynamical coupling between conduction electrons and low-energy phonons can lead to behavior similar to heavy-fermion compounds. Such a scenario is likely relevant for A15 superconductors (Nb$_3$Sn, V$_3$Si), which are well known for their relatively high $T_c$ arising from strong electron-phonon coupling mediated by low-energy phonon modes. However, as will be shown in the superconducting transition analysis that follows, YPt$_2$Si$_2$ exhibits weak electron-phonon coupling. Consequently, the mechanism proposed in Refs.\cite{Yu_1984, Matsuura_1986} cannot account for the enhanced Kadowaki-Woods ratio observed in YPt$_2$Si$_2$, indicating the need for an alternative explanation. 

\begin{figure}[]
\includegraphics[scale=1.4]{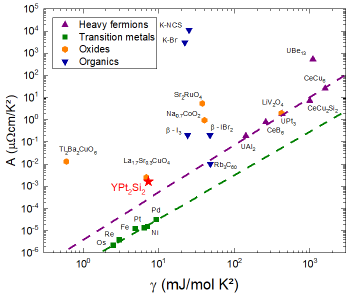}
\caption{\label{fig:KWR} The coefficient of the $T^2$ term, $A$, in $\rho(T)$ below 50 K is plotted as a function of the Sommerfeld coefficient, $\gamma$, obtained from the specific heat for different materials. This representation is widely known as the Kadowaki-Woods (KW) plot \cite{KWR_Jacko2009}. The two dotted straight lines indicate the characteristic trends observed for transition metals and for heavy-fermion compounds respectively. YPt$_2$Si$_2$ falls in the region associated with materials exhibiting strong electronic correlations}
\end{figure}

\begin{table}[]
\scriptsize
\caption{Comparison between the normal state parameters of the material studied in this work, and the published single crystalline LaPt$_2$Si$_2$ \cite{Falkowski_2019_La}.}
\begin{ruledtabular}
\begin{tabular}{lcr}
\label{tab:normal_Params_Y_La}
\textrm{Parameter}&
\textrm{YPt$_2$Si$_2$}&
\textrm{LaPt$_2$Si$_2$}\\
\colrule
T$_{CDW}$ (K)& - & 85\\
$\rho_0$ ($\mu \Omega cm$)& 88.8 & 82.7\\
RRR  & 1.52 & 1.79\\
A ($\mu \Omega cm$/K$^2$) & $1.66(4)\times10^{-3}$ & $3.23\times10^{-3}$\\
$\theta_{BG}$ (K) & 212 & 226 \\
$\theta_D$ (K) & 228(4) & 205 \\
$\gamma$ (mJ/mol$\cdot$ K$^2$)& 7.3(5) & 7.8 \\
A/$\gamma^2$ ($\mu\Omega$cm~(mol-K/mJ)$^2$) & $5.17\times10^{-5}$ & $5.3\times10^{-5}$ \\
Reference & This work & \cite{Falkowski_2019_La} \\ 
\end{tabular}
\end{ruledtabular}
\end{table}

\subsection{Superconducting state}

\begin{figure*}[t]
\centering
\includegraphics[scale=0.45]{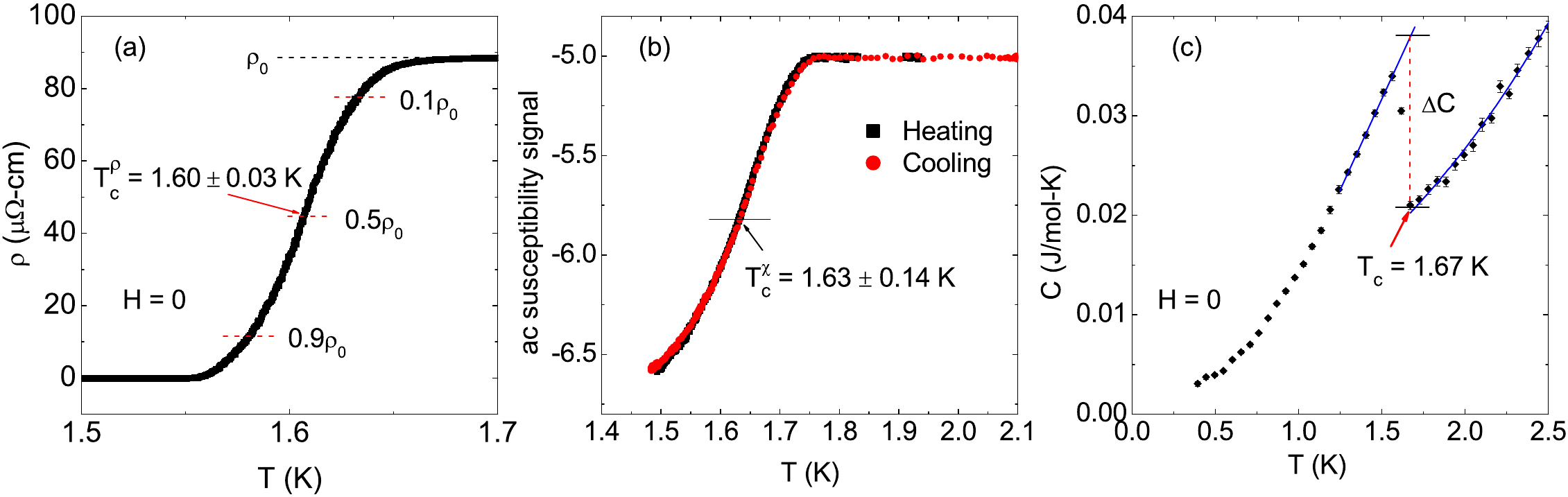}\\
\caption{\label{fig:transitions} Temperature dependence of electrical resistivity, $\rho(T)$, in zero applied magnetic field showing the superconducting transition. $T_c^{\rho}$ is defined as the temperature at which $\rho$ decreases to 50 $\%$ of its normal-state residual value $\rho_0$ (b) Temperature dependence of the ac susceptibility depicting the superconducting transition. $T_c^{\chi}$ is defined as the mid-point of the transition. (c) Temperature dependence of the specific heat, $C(T)$, at $H = 0$, showing a bulk superconducting transition at $T_c \approx 1.67$ K. The method used to estimate the specific heat jump, $\Delta C$ at $T_c$ is indicated in the figure.}
\end{figure*}

The superconducting transition is clearly observed in electrical resistivity, ac magnetic susceptibility and specific heat measurements, as shown in Fig. 5(a-c). The zero-field electrical resistivity exhibits a sharp superconducting transition at $T_c^{\rho}= 1.6$~K with a transition width of 0.06~K, defined by the temperature difference between 10~$\%$ and 90~$\%$ of the residual resistivity, $\rho_0$. The ac-susceptibility data show a clear diamagnetic transition characteristic of superconductivity, with $T_c^{\chi}=1.63$~K, defined as the midpoint of the transition and a broadening of 0.14~K. Specific heat measurements confirm bulk superconductivity with $T_c=1.67$~K, as shown in Fig.~5c. The transition temperatures for our single crystals, obtained from all three measurements, are all consistent within the estimated error and found to be slightly higher than that reported in the literature for polycrystalline samples \cite{SHELTON1984797,Nagano2013}. For parameter estimation, we adopted the bulk $T_c=1.67$~K determined from the specific heat data. The specific heat jump at $T_c$  is $\Delta C =17.14$~mJ/{mol-K}, yielding $\Delta C/\gamma T_c=1.12$ when using $\gamma=9.2(2)$~mJ/mol-K$^{2}$. This value of $\gamma$ differs slightly from that mentioned in the section discussing the normal state, $\gamma=7.3(5)$~mJ/mol-K$^{-2}$, due to the use of a $^3$He cryostat and a separate addenda measurement for the superconducting-state specific heat. Using $\gamma=$~7.3(5)~mJ/mol-K$^{2}$ instead gives $\Delta C/\gamma T_c=1.41$. In both cases, $\Delta C/\gamma T_c$ remains below the BCS weak-coupling limit of 1.43, indicating that YPt$_2$Si$_2$ is a weakly coupled superconductor \cite{tari2003specific}. Moreover, $\Delta C/\gamma T_c$ being less than the BCS weak-coupling limit also suggests the anisotropic or multigap nature of the superconducting gap for YPt$_2$Si$_2$.

\begin{figure}[]
\includegraphics[scale=0.55]{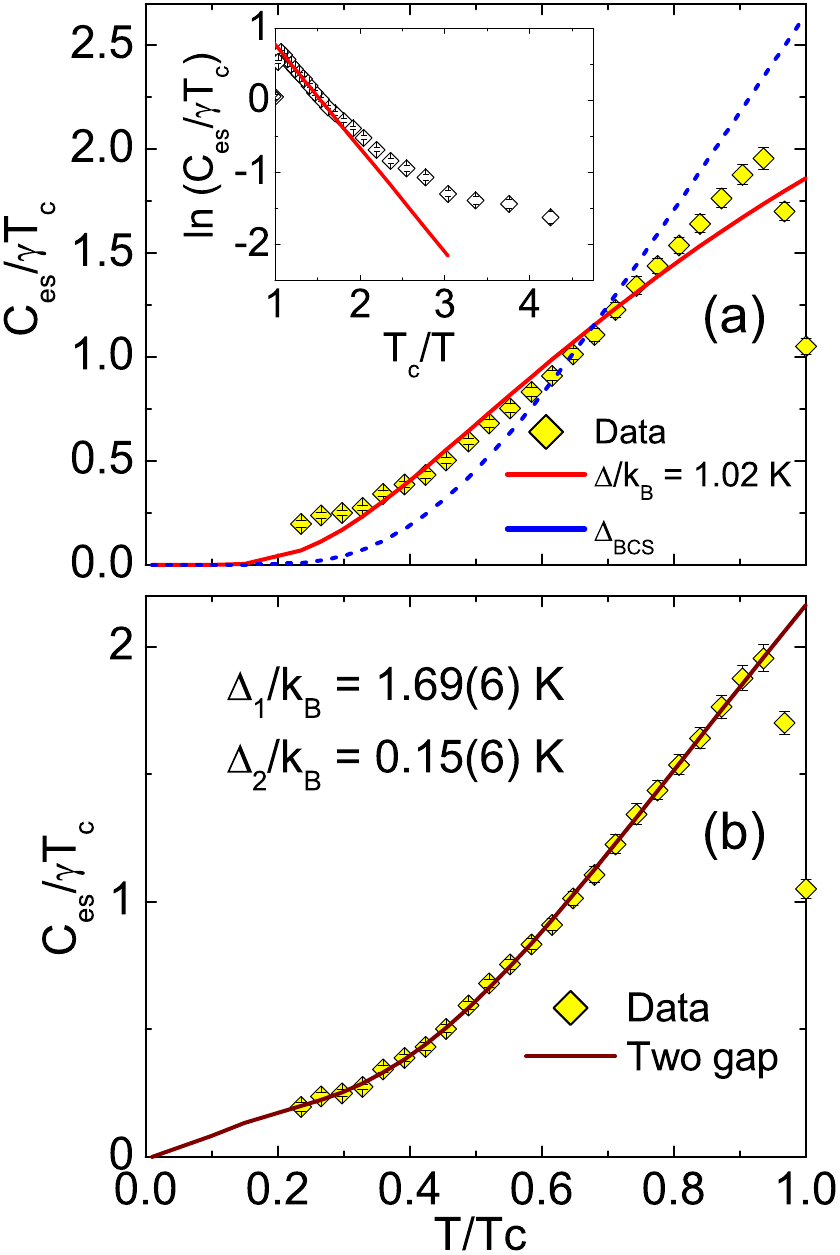}
\caption{\label{fig:SC-gap} (a) Normalized electronic specific heat in the superconducting state, $C_{es}/\gamma T_c$, as a function of the normalized temperature, $T/T_c$. Solid and dashed lines represent fit using Equation 3, with $\Delta/k_B = 1.02$~K, and $\Delta_{\mathrm{BCS}}/k_B = 1.76$~K respectively. Inset shows the ln$(C_{es}/\gamma T_c)$ as a function of $T_c/T$. The solid line displays pronounced nonlinearity at low temperature, clearly deviating from the expected linear behavior of the BCS model. (b) $T/T_c$ dependence of the $C_{es}/\gamma T_c$ in the superconducting state. Data is well explained considering the two-gap model, shown in Equation 4.}
\end{figure}

The electronic specific heat in the superconducting state, $C_{es}$, was obtained by subtracting the phonon contribution, $C_{ph}$, from the total specific heat, i.e., $C_{es} = C_{\mathrm{tot}} - C_{ph}$, where $C_{ph} = \beta T^{3}$. The temperature dependence of $C_{es}$ provides information about the superconducting energy gap and, consequently, the pairing symmetry. For a conventional BCS superconductor, the electronic specific heat in the superconducting state follows an exponential temperature dependence given by \cite{tari2003specific},

\begin{equation}
    C_{es}/\gamma T_c = a~exp{\left(\frac{ - \Delta }{ k_BT }\right)}
\end{equation}
where $a$ is a constant and $\Delta$ is the superconducting energy gap; both are treated as fitting parameters. As shown in Fig. 6(a), this simple BCS expression does not adequately describe the $C_{es}/\gamma T_c$ versus $T/T_c$ data. The fit yields $\Delta/k_B = 1.02(4)$~K, and $a = 5.1(4)$. For comparison, the weak-coupling BCS value $\Delta_{\mathrm{BCS}}/k_B = 1.76$~K also fails to reproduce the experimental behavior. Furthermore, the inset of Fig. 6(a), which presents the logarithm of the normalized electronic specific heat as a function of normalized inverse temperature, shows a clear deviation from the expected linear behavior for an isotropic BCS superconducting gap. 

To achieve a better description of the $C_{es}/\gamma T_c$ vs. $T/T_c$ data, we employed a two-gap model consisting of a weighted sum of two isotropic energy-gaps, given by-

\begin{equation}
    C_{es}/\gamma T_c = b \left[ x~exp{\left(\frac{ - \Delta_1 }{ k_BT }\right)} + (1-x)~exp{\left(\frac{ - \Delta_2 }{ k_BT }\right)}\right],
\end{equation}
where $b$, $\Delta_1/k_B T$, $\Delta_2/k_B T$, and $x$ are fitting parameters.

As shown in Fig.~6(b), the two-gap model reproduces the experimental data well, yielding $b = 10.4(6)$, $\Delta_1/k_B = 1.69(7)$~K, $\Delta_2/k_B = 0.15(6)$~K, and $x = 0.965(6)$. These results indicate that YPt$_2$Si$_2$ is likely a two-gap superconductor with weak electron-phonon coupling, as both gap magnitudes are smaller than the weak-coupling BCS value $\Delta_{\mathrm{BCS}}/k_B = 1.76$~K. Furthermore, since the weighted average of the smaller gap is less than 4~$\%$, the presence of a minimum in the superconducting gap i.e., an anisotropic superconducting gap, cannot be ruled out. This is plausible due to the anisotropic Fermi surface in YPt$_2$Si$_2$ (Fig.~\ref{fig:fs}). To resolve this issue, measurements below 0.2$T_c$ are required. 

The temperature dependence of the electrical resistivity, $\rho(T)$, under different applied magnetic fields, parallel to c-axis, is shown in Fig.~7(a). With increasing magnetic field, the superconducting transition shifts to lower temperatures and exhibits slight transition broadening. For each $\rho(T)$ curve, the superconducting transition temperature at a given magnetic field, $T_c(H)$, was defined as the temperature corresponding to a 50~$\%$ drop from the residual resistivity, $\rho_0$. Using these values, the temperature dependence of the upper critical field, $H_{c2}^{\parallel c}(T)$, was constructed, as shown in Fig. 7(b). The solid line in Fig.~7(b) represents a fit to the Ginzburg-Landau (GL) expression, 

\begin{equation}
    H_{c2}^{\parallel c}(T) = H_{c2}^{\parallel c}(0)\frac{1 - (T/T_c)^2}{1 + (T/T_c)^2},
\end{equation} 
where $H_{c2}^{\parallel c}(0)$ and $T_c$ are fitting parameters corresponding to the zero-temperature upper critical field and the zero-field superconducting transition temperature, respectively. The fitted parameters are $H_{c2}^{\parallel c}(0) \approx 2.74(8)$~kOe, and $T_c = 1.66(3)$~K. The fitted value of $H_{c2}^{\parallel c}(0)$ appears to be slightly underestimated, as the superconducting transition is not fully suppressed at $H = 3$~kOe (see Fig.~7(a)). The obtained $H_{c2}^{\parallel c}(0)$ is found to be comparable to the one reported previously for a polycrystalline sample \cite{Pikul_2017}.  

\begin{figure}[]
\includegraphics[scale=0.55]{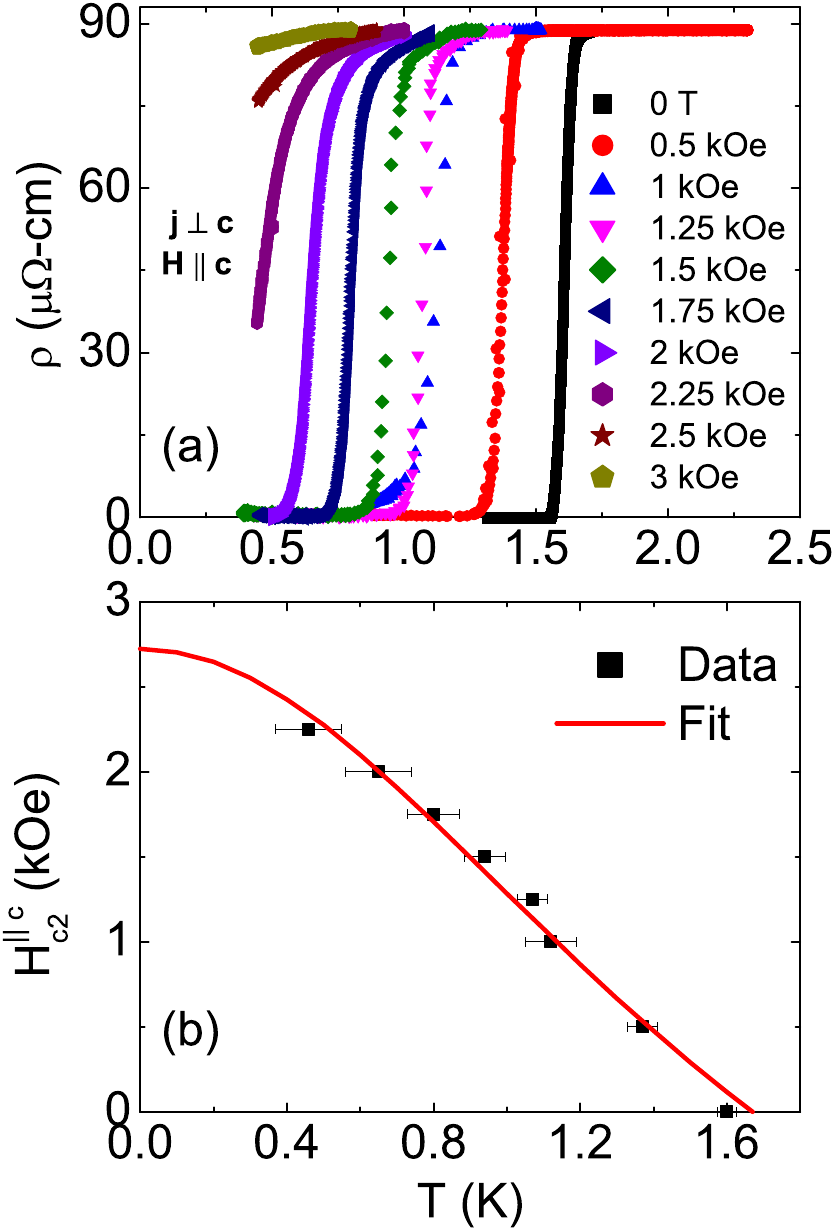}
\caption{\label{fig:Hc2} (a) Electrical resistivity as a function of temperature, $\rho(T)$, measured under different applied magnetic fields, $H$, along the crystallographic $c$-axis. (b) Temperature dependence of the upper critical field, $H_{c2}(T)$, determined from the $\rho(T)$ data shown in the panel (a). The solid line represents a fit using Equation 5. The error bars in the $H_{c2}$ values are estimated from the temperature difference between the 10~$\%$ and 90~$\%$ criteria of $\rho_0$.}
\end{figure}

Using $H_{c2}(0)$, the Ginzburg-Landau coherence length ($\xi$) is determined by,
\begin{equation}
    \xi = \sqrt{\frac{\phi_0}{2\pi H_{c2}(0)}} = 34~nm
\end{equation}

\noindent where $\phi_0~=~2.068\times 10^{-15}~$~Wb is the magnetic flux quantum. Since $l \ll \xi$, the Ginzburg-Landau penetration depth ($\lambda_{GL}$) was calculated using the dirty limit \cite{SC_params_PhysRevB.19.4545}, where 
\begin{equation}
    \lambda_{GL} = 6.42\times 10^{-5} \left(\frac{\rho_0\times 10^{-8}}{T_c}\right)^{1/2}\left(1-\frac{T}{T_c}\right)^{-1/2},
\end{equation}

\noindent giving $\lambda_{GL} = 476$~nm. The resulting Ginzburg-Landau parameter of $\kappa = \lambda_{GL}/\xi = 14$ demonstrates that YPt$_2$Si$_2$ is a type-II superconductor, since $\kappa > 1/\sqrt{2}$ \cite{de1966superconductivity}. The thermodynamic critical field $H_c(0)$ is calculated from
\begin{equation}
    H_c(0) = 4.23 \sqrt{\gamma} T_c,
\end{equation}

\noindent and the obtained value of $H_c(0) = 256~$Oe is used to calculate the lower critical field with the expression
\begin{equation}
    H_{c1}(0) = \frac{H_c ln(\kappa)}{\sqrt{2} \kappa}
\end{equation}
 
This yields a value of $H_{c1}(0) = 34$~Oe. The calculated and experimentally determined superconducting parameters of the YPt$_2$Si$_2$ single crystal are summarized in Table~\ref{tab:Params_Y_La} and compared with those reported for the well-known superconductor LaPt$_2$Si$_2$ and for polycrystalline YPt$_2$Si$_2$ \cite{Falkowski_2019_La, Pikul_2017}. 

\begin{table}[t]
\scriptsize
\caption{Comparison between the material studied in this work, and the published YPt$_2$Si$_2$ (polycrystal \cite{Pikul_2017}) and single crystalline LaPt$_2$Si$_2$ \cite{Falkowski_2019_La}, GL penetration depth ($\lambda_{GL}$), GL coherence length ($\xi$), GL parameter ($\kappa$), lower critical field H$_{c1}$(0), and thermodynamic critical field H$_{c}$ were estimated on dirty limit by the equations described in Ref. \cite{SC_params_PhysRevB.19.4545}.}
\begin{ruledtabular}
\begin{tabular}{lccr}
\label{tab:Params_Y_La}
\textrm{Parameter}&
\textrm{YPt$_2$Si$_2$}&
\textrm{YPt$_2$Si$_2$}&
\textrm{LaPt$_2$Si$_2$}\\
\colrule
Reference & This work & \cite{Pikul_2017} & \cite{Falkowski_2019_La} \\
$T_c$ (K) & 1.67 & 1.54 & 1.6\\
$\frac{\Delta_C}{\gamma T_c}$ & 1.12 & 1.6 & 1.26 \\
$\frac{2\Delta}{K_B T_c}$ & - & - & 2.73 \\
$\lambda_{ep}$ & 0.51 & 0.50 & 0.53 \\
$N(E_f)$ (states/eV/f.u.) & 2.05 & 2.58& 2.26 \\
$\mu_0 H_{c2}(0)~$(Oe) (GL equation)& 2843 & 2500 & 1921 \\
$\xi~$(\AA) (GL equation)  & 340 & 320 & 413.8 \\
$\lambda_{GL}$ (\AA)& 4763 & 2530 & 4610 \\
$\kappa$ & 14 & 7 & 23.5 \\
$H_c$ (Oe)& 256 & 250 & 257 \\
$H_{c1}$(0) (Oe)& 34 & 50 & 24 \\
\end{tabular}
\end{ruledtabular}
\end{table}

The electron-phonon coupling constant ($\lambda_{ep}$) was estimated by employing $\theta_D$ and $T_c$ in the expression given by McMillan \cite{Mcmillan_1968},  

\begin{equation}
    \lambda_{ep} = \frac{1.04 + \mu^* ln\left(\frac{\theta_D}{1.45 T_c}\right)}{(1-0.62\mu^* )ln\left(\frac{\theta_D}{1.45 T_c}\right) -1.04}
\end{equation}

\noindent where $\mu^*$ is the Coulomb pseudopotential considered 0.13 for transition metals, resulting in $\lambda_{ep} = 0.51$, similar to LaPt$_2$Si$_2$ \cite{Falkowski_2019_La}, confirming the weak coupling strength. 

\section{Theoretical results}

A non-magnetic ground state was obtained for this compound, as expected for this class of superconducting materials. The calculated crystallographic properties show that YPt$_2$Si$_2$ possesses a tetragonal crystal structure (Fig.~\ref{fig:structure}(b)) with lattice parameters $a = 4.1615$~\AA, $c = 9.7366$~\AA, unit cell volume $V = 168.62$~\AA$^3$, and bulk modulus $B = 170.23$~GPa (obtained by fitting the calculated volumes to a Birch--Murnaghan equation of state). Comparison with the experimental values in Table~\ref{tab:XRD_parameters} reveals that the absolute errors for the lattice constants and volume are below 1\%, indicating a good description of the compound's structure.

The mechanical properties of a material are intrinsically linked to its electronic structure and bonding characteristics, which can, in turn, influence its superconducting behavior. The elastic constants describe a material's response to external stresses and provide insights into the interatomic bond strength. We investigated the mechanical properties of YPt$_2$Si$_2$ by calculating its six independent elastic constants~\cite{PhysRevB.72.035105, kresse1999ultrasoft,hobbs2000fully} for the tetragonal structure: $C_{11}$, $C_{12}$, $C_{13}$, $C_{33}$, $C_{44}$, and $C_{66}$. The calculated values are 240.7, 97.3, 150.9, 247.1, 14.1, and 67.9~GPa, respectively. These values indicate high resistance to compression along the $a$ and $b$ axes (equal in length for this tetragonal structure, $C_{11}$), and the $c$ axis ($C_{33}$), but low resistance to shear deformation involving the $bc$ plane ($C_{44}$).

These constants must satisfy the following mechanical stability criteria~\cite{djied2014structural,wu2007crystal}:

\begin{equation*}
\begin{aligned}
C_{11} &> 0, \\
C_{33} &> 0, \\
C_{44} &> 0, \\
C_{66} &> 0, \\
C_{11} - C_{12} &> 0, \\
C_{11} + C_{33} - 2C_{13} &> 0, \\
2C_{11} + C_{33} + 2C_{12} + 4C_{13} &> 0,
\end{aligned}
\end{equation*}
the calculated elastic constants satisfy these criteria for a tetragonal structure, confirming its mechanical stability. Additionally, the dynamical stability is guaranteed by the absence of imaginary frequencies in the calculated phonon dispersion relation at the equilibrium volume (Fig.~\ref{fig:fonons}).

The Debye temperature, $\theta_D$, which is related to the average sound velocity and atomic concentration, is a measure of material stiffness and an important parameter in superconductivity theory as it connects to the maximum phonon frequency contributing to electron-phonon interaction. $\theta_D$ can be obtained from the elastic moduli using:

\begin{equation}
\theta_D = \frac{\hbar s q_D}{k_B},
\end{equation}
where $s = \left(\frac{1}{3s_l^3} + \frac{2}{3s_t^3}\right)^{-1/3}$ is the average sound velocity, $s_l = \sqrt{\frac{B + \frac{4}{3}G}{\rho}}$ is the longitudinal sound velocity, $s_t = \sqrt{\frac{G}{\rho}}$ is the transverse sound velocity, $G$ is the isotropic shear modulus (calculated from the crystalline lattice constants, see supplemental information), $q_D = \sqrt[3]{6\pi^2\eta_\alpha}$, and $\eta_\alpha$ is the atomic concentration. The calculated Debye temperature for YPt$_2$Si$_2$ is 264.4~K, in reasonable agreement with the experimental value of 228(4)~K (an error of 13.8\%). This difference may arise from anharmonic effects not included in our calculations and the presence of defects in experimental samples.

\begin{figure}[]
\includegraphics[scale=0.45]{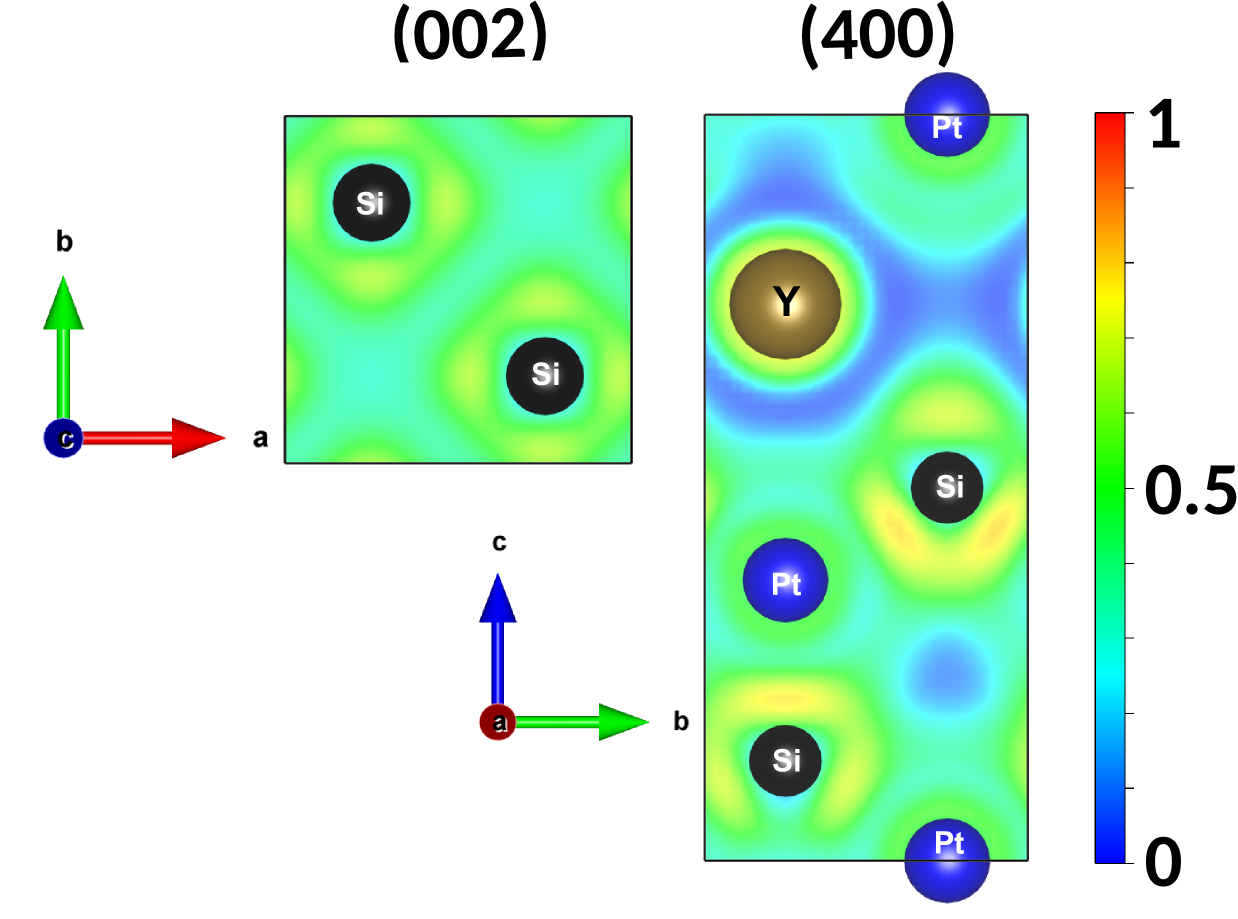}
\caption{\label{fig:elf} Visualization of the electron localization function (ELF) in the (002) and (400) planes of YPt$_2$Si$_2$.}
\end{figure}

To investigate the bonding properties of YPt$_2$Si$_2$, we analyzed the electron localization function (ELF), a measure of the probability of finding an electron in the vicinity of a reference electron with the same spin. ELF is commonly used to visualize chemical bonding in molecules and solids, based on the principle that electrons are less likely to be found in highly localized regions. An ELF value of 1 indicates perfect localization, while 0.5 
corresponds to a homogeneous electron gas~\cite{becke1990simple,savin1997elf}. Ionic bonding is characterized by high ELF values near the nuclei and very low values ($\sim$0) in the interstitial region. Covalent bonding between two atoms shows a local ELF maximum along the bond axis, typically ranging from 0.6 to 1.0, correlating with bond strength. Metallic bonding represents an intermediate case with a relatively uniform ELF distribution in the interstitial region, typically between 0.3 and 0.6.

Figure~\ref{fig:elf} displays ELF sections in the (002) and (400) planes. The (002) plane, defined by Si1 and Si2 atoms forming zigzag chains along the [100] and [010] directions, exhibits a metallic-ionic bond with a maximum ELF value of 0.36. The (400) plane, representing a YPt$_2$Si$_2$ layer where Y atoms form chains along the [010] direction, shows an ELF value close to zero around Y atoms, characteristic of ionic bonding. The Y-Pt2 bond has an ELF value around 0.35, indicating a metallic-ionic character, while the Y-Si2 bond with an ELF value of 0.46 suggests metallic bonding. Finally, the Si2-Pt2 bond is characterized by a metallic-covalent interaction with an ELF value of approximately 0.7.

\begin{figure}[]
\includegraphics[scale=0.7]{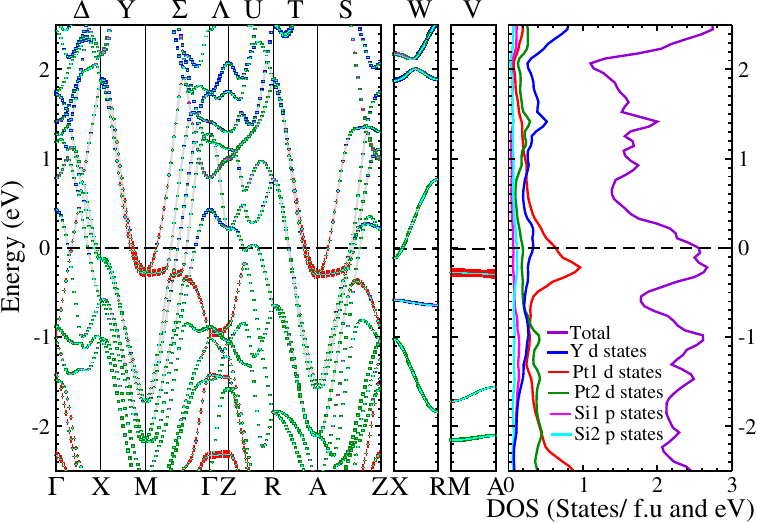}
\caption{\label{fig:bandas} Left panel: Electronic structure of YPt$_2$Si$_2$ including spin-orbit coupling. Fat band dispersion showing the angular contribution of the atomic species: Y (blue), Pt1 (red), Pt2 (green), Si1 (magenta), and Si2 (cyan). The five conduction bands that cross the Fermi level (dashed line) are highlighted with gray shading. Far-right panel: Total and projected density of states (DOS), showing the contributions from each atomic species.}
\end{figure}

The electronic properties of YPt$_2$Si$_2$ were investigated by calculating the band dispersion along high-symmetry directions of the first Brillouin zone (FBZ), along with the total and projected density of states (DOS). The band dispersion and DOS near the Fermi level are presented in Fig.~\ref{fig:bandas}. Five bands cross the Fermi level, confirming the metallic nature of this compound. Topology analysis reveals a hole-like character for the lowest conduction band along the $\Sigma$, $\Lambda$, and S directions, transitioning to electron-like along the $\Delta$ and U directions. The two intermediate conduction bands exhibit both hole-like (along $\Sigma$, T, and S) and electron-like (along $\Delta$ and W) characters. Finally, the highest two conduction bands show only electron-like character along the Y, $\Sigma$, T, and S directions. Notably, no bands cross the Fermi level along the V direction, in contrast to the 
$\Sigma$ and S directions, which are crossed by all five conduction bands.

The fat band dispersion and projected DOS highlight the dominant angular contributions. Pt1 $d$ states (red) are prevalent in the highest two conduction bands along the Y, $\Sigma$, T, and S directions. Pt2 $d$ states (green) contribute significantly to the intermediate bands along the $\Delta$, $\Sigma$, U, T, S, and W directions, consistent with previous findings~\cite{kim_2015}. Y $d$ states (blue) primarily contribute to the lowest conduction band along the $\Delta$, $\Sigma$, U, S, and W directions.

Furthermore, the lower valence bands (not shown here) are composed of Si1 and Si2 $s$ and $p$ states and Pt $d$ states in the energy range of -12 to -7.8 eV. The intermediate valence band (between -7.6 and -3.8 eV) is mainly derived from Pt $d$ states with a minor contribution from Pt $s$ states. The upper valence bands are predominantly formed by Pt $d$ states with a small admixture of Y $d$ states.

\begin{figure*}[t]
\includegraphics[scale=0.55]{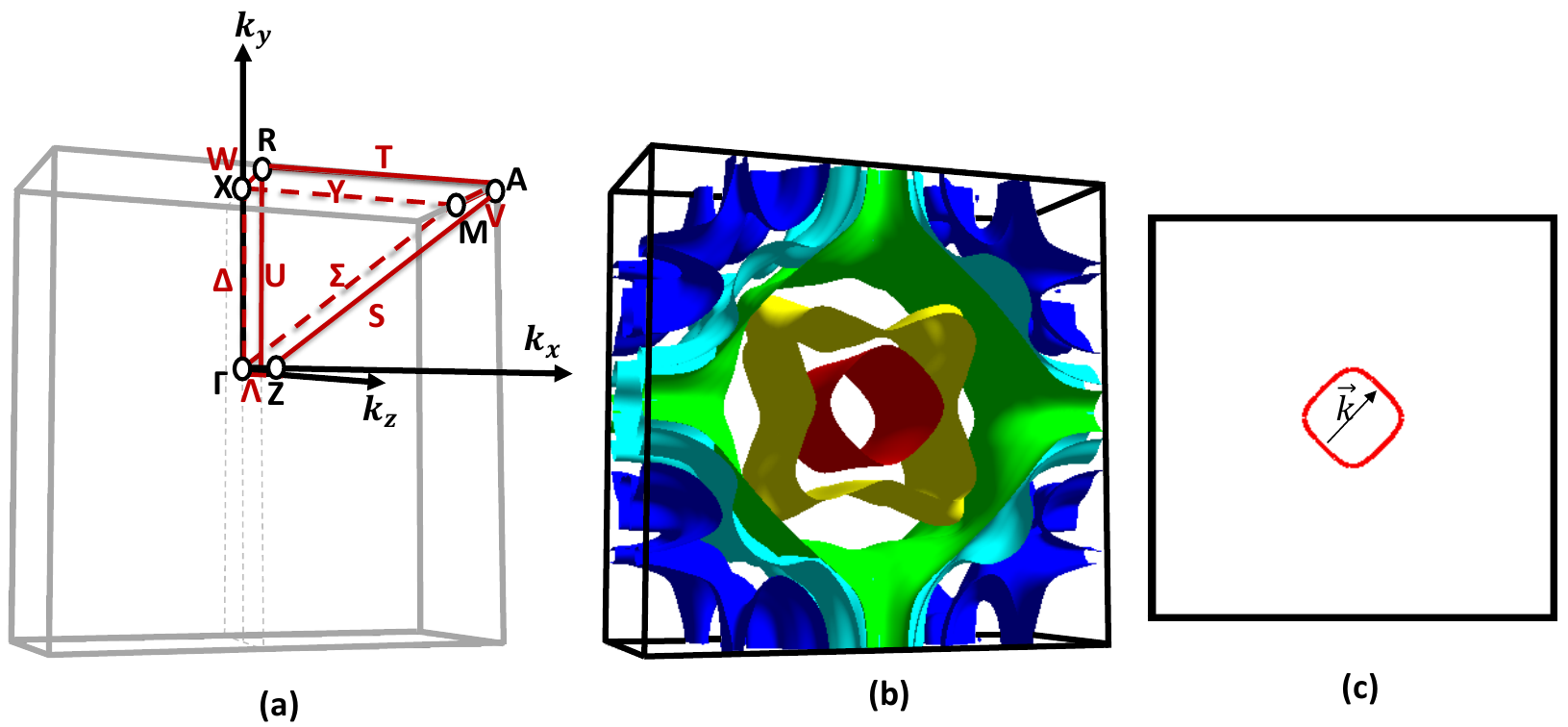}
\caption{\label{fig:fs} (a) Primitive Brillouin zone (PBZ) of YPt$_2$Si$_2$ (space group $P4/nmm$) illustrating the high-symmetry points (black) and directions (red) used in the calculations. (b) Calculated Fermi surface of YPt$_2$Si$_2$ within the reciprocal primitive cell. (c) Nesting Fermi surface.}
\end{figure*}

The Fermi surface (FS), depicted in Fig.~\ref{fig:fs} alongside the first Brillouin zone (FBZ) and the nesting vector, exhibits significant anisotropy across its five distinct sheets. The first branch (blue surface) forms a central cylinder oriented along the V direction, featuring four stacked egg-shaped features within it. Notably, this branch connects the Brillouin zone edges solely along the $k_z$ axis. The second (cyan) and third (green) branches, primarily derived from Y and Pt2 $d$ states, form a network of interconnected cylinders with their principal axis aligned along the $\Lambda$ direction, extending throughout the entire Brillouin zone along all crystallographic axes. In contrast, the fourth (yellow) and fifth (red) branches are also cylindrical with their principal axis along the $\Lambda$ direction, but their extension within the Brillouin zone is limited to this direction. These branches are predominantly composed of Pt1 $d$ states. Interestingly, the fifth branch displays parallel sections connected by a single wave vector corresponding to half of a reciprocal lattice vector, indicating intra-branch FS nesting as illustrated in Fig.~\ref{fig:fs}(c). Such intra-branch nesting favors electron pairing within the same conduction band, potentially leading to single-band superconductivity analogous to the BCS mechanism.

\begin{figure*}[t]
  \includegraphics[scale=0.5]{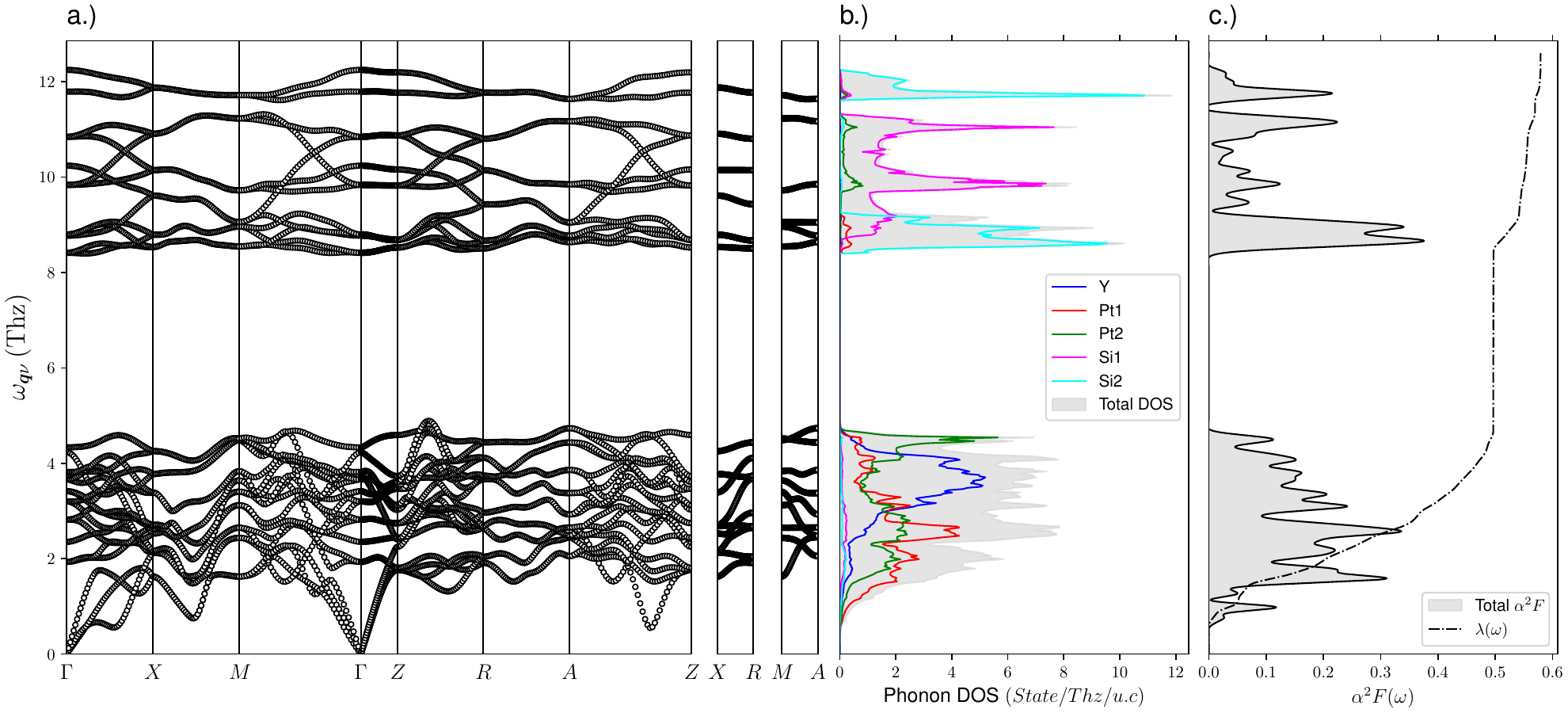}
  \caption{\label{fig:fonons}
  Calculated phonon dispersion, phonon density of states, and Eliashberg spectral function of YPt$_2$Si$_2$.
  (a) Phonon dispersion along high-symmetry directions of the Brillouin zone.
  (b) Total phonon DOS (gray shading) and atom-projected DOS: Y (blue), Pt1 (red), Pt2 (green), Si1 (magenta), and Si2 (cyan).
  (c) Total Eliashberg spectral function $\alpha^2F(\omega)$ (gray shading) together with the cumulative electron–phonon coupling
  $\lambda_{ep}(\omega)=2\int_0^\omega d\omega' \alpha^2F(\omega')/\omega'$ .}
\end{figure*}

Figure \ref{fig:fonons} presents the phonon dispersion relations, the total and atom-resolved phonon density of states $F(\omega)$, and the Eliashberg spectral function $\alpha^{2}F(\omega)$, all calculated without including SOC.
The primitive cell of these systems contains ten atoms, resulting in a phonon spectrum with 3 acoustic and 27 optical branches. This spectrum can be divided into four frequency regions: acoustic (0-2.2~THz), low optic (2.2-3.5~THz), intermediate optic (3.5-4.7~THz), and high optic (8.3-12.1~THz). The acoustic (3 bands) and low optic branches (5 bands) are dominated by phonon modes involving Pt1 and Pt2 atomic displacements. The significant mass difference between
Pt atoms ($m_{\text{Pt}} = 195.084 \, \text{amu}$) and Y ($m_{\text{Y}} = 88.906 \, \text{amu}$) and Si ($m_{\text{Si}} = 28.0855 \, \text{amu}$) atoms leads to a substantial frequency gap of approximately 3.6~THz, separating the intermediate optic (10 bands) and high optic modes (12 bands). Intermediate optic modes exhibit contributions from both Y and Pt atomic displacements, whereas the high optic modes show an appreciable contribution from Si atomic displacements. The intermediate and high frequency regions are separated by a phonon gap of approximately 3.7~THz. A second phonon gap of 0.35~THz is observed within the high optic modes above 11.1~THz.

The Eliashberg spectral function $\alpha^2F(\omega)$ displays a peak structure broadly similar to the total phonon density of states. Dominant peaks are located at 8.66 THz and 8.95 THz, and are associated with Si2 phonon modes. Additional prominent peaks appear at lower frequencies, notably at 2.57 and 1.58 THz, which are dominated by Pt phonon modes, and at 3.10 THz, where Pt and Y derived contributions are comparable. The cumulative coupling $\lambda_{ep}(\omega)$ increases rapidly over the 0-4.8 THz range, reaching $\sim 80 \%$ of its final value, while the Si dominated high frequency modes provide a smaller contribution. Consistently, the obtained logarithmic average phonon 
frequency is $\omega_{\ln}=2.65$ THz, indicating that the effective pairing interaction is governed primarily by phonons in the low optical frequency section of the spectrum dominated by Pt and Y related phonon modes. Together with the electronic-structure results, which show that the bands dispersing across the BZ have substantial Pt2 and Y $d$-state character, these findings suggest that superconductivity in this compound is primarily associated with Pt and Y $d$-states coupled to low-energy phonons. For all subsequent calculations in which the Brillouin-zone integrals containing products of Dirac delta functions are evaluated, we employed a Gaussian $\sigma = 0.1$ Ry, which yields $\lambda_{ep} = 0.58$ and is the smallest broadening that keeps $\lambda_{ep}$ converged within $4\times 10^{-3}$.  Using the Allen-Dynes modified McMillan formula, we obtained a critical temperature $T_c = 1.8$ K. These values are in reasonable agreement with our experimental findings and with the polycrystal samples reported in Ref \cite{Pikul_2017} ($\lambda_{ep} \sim 0.6$ and $T_c \sim 1.5$ K). The two approaches regards  the weak coupling regime of the superconductivity, where all values are below the BCS limit, the similarities between the superconducting behavior and the electronic structure of YPt$_2$Si$_2$ and LaPt$_2$Si$_2$ indicates the possibility of a similar gap structure \cite{nie_2021}.

In some materials, charge density wave (CDW) formation is driven by electron-phonon coupling, where lattice and electronic instabilities are coupled, resulting in the CDW wave vector $q_{CDW}$ vanishing at the transition temperature $T_{CDW}$, as seen in NbSe$_2$. In others, CDWs are driven by Fermi surface nesting, with lattice distortion being a secondary effect~\cite{class_cdw,CDW_thorne_1996}. Materials with the CaBe$_2$Si$_2$-like structure fall into the latter category. For these materials, a crystal lattice volume threshold of approximately 170~\AA$^3$ can be inferred, as smaller compounds like YPt$_2$Si$_2$ and LuPt$_2$Si$_2$ do not exhibit CDWs. In contrast, larger compounds such as UPt$_2$Si$_2$ \cite{UPt2Si2_2023}, LaPt$_2$Si$_2$, SrPt$_2$As$_2$\cite{kim_2015, SrPt2As2_2010}, NdPt$_2$Si$_2$\cite{NdPt2Si2_2020}, and PrPt$_2$Si$_2$\cite{Kumar2010} show CDWs alongside lattice distortions, despite having weak electron-phonon coupling.

The Fermi surface of YPt$_2$Si$_2$ exhibits nesting similar to that of SrPt$_2$As$_2$, which leads to a high density of electronic states with comparable energies and momenta, enhancing the likelihood of electron-phonon interactions. However, experimental observation of CDW formation in YPt$_2$Si$_2$ is absent, possibly due to a large interatomic distance in the $c$-plane and a reduction in the density of states at the Fermi level~\cite{kim_2015}. This is consistent with the understanding that Fermi surface nesting alone is insufficient to drive CDW formation without a strong wave vector dependence of the electron-phonon coupling~\cite{La_syncrotron_NOCERINO2023100621}.

\section{Summary}

In summary, we successfully synthesized high-quality single crystals of YPt$_2$Si$_2$ using the Sn-flux method. The normal-state resistivity $\rho(T)$ in the temperature range 2-300 K does not follow the Bloch-Gr\"{u}neisen law. Instead, $\rho(T)$ exhibits an unusual linear temperature dependence between 50 and 300 K and a quadratic temperature dependence below 50 K. Such an extended linear behavior is uncommon in the presence of weak electron-phonon coupling and warrants further investigation. Despite a relatively small Sommerfeld coefficient, a large Kadowaki-Woods ratio $A/\gamma^2 = 5.17 \times 10^{-5}$ $\mu\ohm$-cm~(mol-K/mJ)$^2$ is obtained, a value typically associated with strong electron-electron correlations. These results suggest that YPt$_2$Si$_2$ is not a simple metallic system and the enhanced electronic correlations may possibly have its origin different than the one for conventional correlated electron system. A sharp superconducting transition is observed in electrical resistivity, ac susceptibility, and specific heat measurements, confirming bulk superconductivity with $T_c = 1.67$ K. The specific heat jump satisfies $\Delta C/\gamma T_c < 1.43$, consistent with weak electron-phonon coupling. The temperature dependence of the  specific heat in the superconducting-state is well described by a two-gap model with two isotropic gaps. In addition, the positive curvature of $H_{c2}^{\parallel c}(T)$ near $T_c$ further supports multigap superconductivity in YPt$_2$Si$_2$. First-principles DFT calculations complement the experimental results by providing detailed insight into the electronic structure, Fermi surface, density of states, phonon spectrum, and electron localization function. Moreover, the coupling between Pt1 atomic vibrations and its $d$ electrons is likely responsible for the superconductivity in this material.

Further studies are being carried out using $\mu$SR techniques to investigate the pairing symmetry of the Cooper pairs, the superconducting gap structure, and search for the spontaneous magnetization that arises from the breaking of time-reversal symmetry (TRS) in noncentrosymmetric superconductors.

\section{Acknowledgments} We acknowledge the financial support from Brazilian funding agencies CAPES, CNPq (Contract Nos. 140921/2022-2, 88887.837417/2023-00, 180596/2025-0), FAPESP (No. 2017/20989-8, No. 2017/10581-1), Colombian agency COLCIENCIAS Convocatoria Doctorados Nacionales No. 757 de 2016), and Vicerrectoría de Investigación Universidad de Antioquia Estrategia de Sostenibilidad No. 2018-2019 (Colombia). ANID Postdoctoral Research Associate, 3250210 (Chile). We acknowledge the computational resources provided by the supercomputing infrastructure of the NLHPC (CCSS210001). The measurements with PPMS with a $^3$He insert were performed at N-BARD, Hiroshima University, Japan. Part of this work was financially supported by a Grant-in-Aid from MEXT/JSPS of Japan [Grant No. JP24K00574] and by a Grant-in-Aid for Transformative Research Areas (A) “Asymmetric Quantum Matters”, JSPS KAKENHI Grant No. JP23H04870. We also acknowledge the experimental support from Multiuser Central Facilities (UFABC), LCCEM (UFABC) and LNNano under the proposals No. 20230920 and No. 20240535. MAA acknowledges the support of the INCT project Advanced Quantum Materials, involving the Brazilian agencies CNPq (Proc. 408766/2024-7), FAPESP (Proc. 2025/27091-3), and CAPES. We thank Dr. Henrique Fabrelli Ferreira for the help with the single crystal Laue diffraction measurement.


\bibliographystyle{apsrev}
\bibliography{apssamp}

\begin{thebibliography}{54}
\expandafter\ifx\csname natexlab\endcsname\relax\def\natexlab#1{#1}\fi
\expandafter\ifx\csname bibnamefont\endcsname\relax
  \def\bibnamefont#1{#1}\fi
\expandafter\ifx\csname bibfnamefont\endcsname\relax
  \def\bibfnamefont#1{#1}\fi
\expandafter\ifx\csname citenamefont\endcsname\relax
  \def\citenamefont#1{#1}\fi
\expandafter\ifx\csname url\endcsname\relax
  \def\url#1{\texttt{#1}}\fi
\expandafter\ifx\csname urlprefix\endcsname\relax\def\urlprefix{URL }\fi
\providecommand{\bibinfo}[2]{#2}
\providecommand{\eprint}[2][]{\url{#2}}

\bibitem[{\citenamefont{Bauer et~al.}(2004)\citenamefont{Bauer, Hilscher,
  Michor, Paul, Scheidt, Gribanov, Seropegin, No\"el, Sigrist, and
  Rogl}}]{CePt3Si_2004}
\bibinfo{author}{\bibfnamefont{E.}~\bibnamefont{Bauer}},
  \bibinfo{author}{\bibfnamefont{G.}~\bibnamefont{Hilscher}},
  \bibinfo{author}{\bibfnamefont{H.}~\bibnamefont{Michor}},
  \bibinfo{author}{\bibfnamefont{C.}~\bibnamefont{Paul}},
  \bibinfo{author}{\bibfnamefont{E.~W.} \bibnamefont{Scheidt}},
  \bibinfo{author}{\bibfnamefont{A.}~\bibnamefont{Gribanov}},
  \bibinfo{author}{\bibfnamefont{Y.}~\bibnamefont{Seropegin}},
  \bibinfo{author}{\bibfnamefont{H.}~\bibnamefont{No\"el}},
  \bibinfo{author}{\bibfnamefont{M.}~\bibnamefont{Sigrist}}, \bibnamefont{and}
  \bibinfo{author}{\bibfnamefont{P.}~\bibnamefont{Rogl}},
  \bibinfo{journal}{Phys. Rev. Lett.} \textbf{\bibinfo{volume}{92}},
  \bibinfo{pages}{027003} (\bibinfo{year}{2004}),
  \urlprefix\url{https://link.aps.org/doi/10.1103/PhysRevLett.92.027003}.

\bibitem[{\citenamefont{Naskar et~al.}(2021)\citenamefont{Naskar, Mishra, Ash,
  and Ganguli}}]{Naskar2021}
\bibinfo{author}{\bibfnamefont{M.}~\bibnamefont{Naskar}},
  \bibinfo{author}{\bibfnamefont{P.~K.} \bibnamefont{Mishra}},
  \bibinfo{author}{\bibfnamefont{S.}~\bibnamefont{Ash}}, \bibnamefont{and}
  \bibinfo{author}{\bibfnamefont{A.~K.} \bibnamefont{Ganguli}},
  \bibinfo{journal}{Bulletin of Materials Science}
  \textbf{\bibinfo{volume}{44}}, \bibinfo{pages}{278} (\bibinfo{year}{2021}),
  ISSN \bibinfo{issn}{0973-7669},
  \urlprefix\url{https://doi.org/10.1007/s12034-021-02587-z}.

\bibitem[{\citenamefont{Smidman et~al.}(2017)\citenamefont{Smidman, Salamon,
  Yuan, and Agterberg}}]{Smidman_2017}
\bibinfo{author}{\bibfnamefont{M.}~\bibnamefont{Smidman}},
  \bibinfo{author}{\bibfnamefont{M.~B.} \bibnamefont{Salamon}},
  \bibinfo{author}{\bibfnamefont{H.~Q.} \bibnamefont{Yuan}}, \bibnamefont{and}
  \bibinfo{author}{\bibfnamefont{D.~F.} \bibnamefont{Agterberg}},
  \bibinfo{journal}{Reports on Progress in Physics}
  \textbf{\bibinfo{volume}{80}}, \bibinfo{pages}{036501}
  (\bibinfo{year}{2017}),
  \urlprefix\url{https://dx.doi.org/10.1088/1361-6633/80/3/036501}.

\bibitem[{\citenamefont{Bauer and Sigrist}(2012)}]{bauer2012non}
\bibinfo{author}{\bibfnamefont{E.}~\bibnamefont{Bauer}} \bibnamefont{and}
  \bibinfo{author}{\bibfnamefont{M.}~\bibnamefont{Sigrist}},
  \emph{\bibinfo{title}{Non-Centrosymmetric Superconductors: Introduction and
  Overview}}, Lecture Notes in Physics (\bibinfo{publisher}{Springer Berlin
  Heidelberg}, \bibinfo{year}{2012}), ISBN \bibinfo{isbn}{9783642246241},
  \urlprefix\url{https://books.google.com.br/books?id=nDZ4lKD00t8C}.

\bibitem[{\citenamefont{Bhattacharyya et~al.}(2019)\citenamefont{Bhattacharyya,
  Adroja, Panda, Saha, Das, Machado, Cigarroa, Grant, Fisk, Hillier
  et~al.}}]{ThCoC2_2019}
\bibinfo{author}{\bibfnamefont{A.}~\bibnamefont{Bhattacharyya}},
  \bibinfo{author}{\bibfnamefont{D.~T.} \bibnamefont{Adroja}},
  \bibinfo{author}{\bibfnamefont{K.}~\bibnamefont{Panda}},
  \bibinfo{author}{\bibfnamefont{S.}~\bibnamefont{Saha}},
  \bibinfo{author}{\bibfnamefont{T.}~\bibnamefont{Das}},
  \bibinfo{author}{\bibfnamefont{A.~J.~S.} \bibnamefont{Machado}},
  \bibinfo{author}{\bibfnamefont{O.~V.} \bibnamefont{Cigarroa}},
  \bibinfo{author}{\bibfnamefont{T.~W.} \bibnamefont{Grant}},
  \bibinfo{author}{\bibfnamefont{Z.}~\bibnamefont{Fisk}},
  \bibinfo{author}{\bibfnamefont{A.~D.} \bibnamefont{Hillier}},
  \bibnamefont{et~al.}, \bibinfo{journal}{Phys. Rev. Lett.}
  \textbf{\bibinfo{volume}{122}}, \bibinfo{pages}{147001}
  (\bibinfo{year}{2019}),
  \urlprefix\url{https://link.aps.org/doi/10.1103/PhysRevLett.122.147001}.

\bibitem[{\citenamefont{Shang et~al.}(2022)\citenamefont{Shang, Ghosh, Smidman,
  Gawryluk, Baines, Wang, Xie, Chen, Ajeesh, Nicklas
  et~al.}}]{LaPtSi_Shang2022}
\bibinfo{author}{\bibfnamefont{T.}~\bibnamefont{Shang}},
  \bibinfo{author}{\bibfnamefont{S.~K.} \bibnamefont{Ghosh}},
  \bibinfo{author}{\bibfnamefont{M.}~\bibnamefont{Smidman}},
  \bibinfo{author}{\bibfnamefont{D.~J.} \bibnamefont{Gawryluk}},
  \bibinfo{author}{\bibfnamefont{C.}~\bibnamefont{Baines}},
  \bibinfo{author}{\bibfnamefont{A.}~\bibnamefont{Wang}},
  \bibinfo{author}{\bibfnamefont{W.}~\bibnamefont{Xie}},
  \bibinfo{author}{\bibfnamefont{Y.}~\bibnamefont{Chen}},
  \bibinfo{author}{\bibfnamefont{M.~O.} \bibnamefont{Ajeesh}},
  \bibinfo{author}{\bibfnamefont{M.}~\bibnamefont{Nicklas}},
  \bibnamefont{et~al.}, \bibinfo{journal}{npj Quantum Materials}
  \textbf{\bibinfo{volume}{7}}, \bibinfo{pages}{35} (\bibinfo{year}{2022}),
  ISSN \bibinfo{issn}{2397-4648},
  \urlprefix\url{https://doi.org/10.1038/s41535-022-00442-w}.

\bibitem[{\citenamefont{Sundar et~al.}(2021)\citenamefont{Sundar, Dunsiger,
  Gheidi, Akella, C\^ot\'e, \"Ozdemir, Lee-Hone, Broun, Mun, Honda
  et~al.}}]{Sundar2021}
\bibinfo{author}{\bibfnamefont{S.}~\bibnamefont{Sundar}},
  \bibinfo{author}{\bibfnamefont{S.~R.} \bibnamefont{Dunsiger}},
  \bibinfo{author}{\bibfnamefont{S.}~\bibnamefont{Gheidi}},
  \bibinfo{author}{\bibfnamefont{K.~S.} \bibnamefont{Akella}},
  \bibinfo{author}{\bibfnamefont{A.~M.} \bibnamefont{C\^ot\'e}},
  \bibinfo{author}{\bibfnamefont{H.~U.} \bibnamefont{\"Ozdemir}},
  \bibinfo{author}{\bibfnamefont{N.~R.} \bibnamefont{Lee-Hone}},
  \bibinfo{author}{\bibfnamefont{D.~M.} \bibnamefont{Broun}},
  \bibinfo{author}{\bibfnamefont{E.}~\bibnamefont{Mun}},
  \bibinfo{author}{\bibfnamefont{F.}~\bibnamefont{Honda}},
  \bibnamefont{et~al.}, \bibinfo{journal}{Phys. Rev. B}
  \textbf{\bibinfo{volume}{103}}, \bibinfo{pages}{014511}
  (\bibinfo{year}{2021}),
  \urlprefix\url{https://link.aps.org/doi/10.1103/PhysRevB.103.014511}.

\bibitem[{\citenamefont{Szytuła and
  Leciejewicz}(1989)}]{SZYTULA1989133_compendio}
\bibinfo{author}{\bibfnamefont{A.}~\bibnamefont{Szytuła}} \bibnamefont{and}
  \bibinfo{author}{\bibfnamefont{J.}~\bibnamefont{Leciejewicz}}
  (\bibinfo{publisher}{Elsevier}, \bibinfo{year}{1989}),
  vol.~\bibinfo{volume}{12} of \emph{\bibinfo{series}{Handbook on the Physics
  and Chemistry of Rare Earths}}, pp. \bibinfo{pages}{133--211},
  \urlprefix\url{https://www.sciencedirect.com/science/article/pii/S0168127389120078}.

\bibitem[{\citenamefont{Shelton et~al.}(1984)\citenamefont{Shelton, Braun, and
  Musick}}]{SHELTON1984797}
\bibinfo{author}{\bibfnamefont{R.}~\bibnamefont{Shelton}},
  \bibinfo{author}{\bibfnamefont{H.}~\bibnamefont{Braun}}, \bibnamefont{and}
  \bibinfo{author}{\bibfnamefont{E.}~\bibnamefont{Musick}},
  \bibinfo{journal}{Solid State Communications} \textbf{\bibinfo{volume}{52}},
  \bibinfo{pages}{797} (\bibinfo{year}{1984}), ISSN \bibinfo{issn}{0038-1098},
  \urlprefix\url{https://www.sciencedirect.com/science/article/pii/0038109884900085}.

\bibitem[{\citenamefont{Vali\v{s}ka et~al.}(2012)\citenamefont{Vali\v{s}ka,
  Posp\'{\i}\v{s}il, Prokle\v{s}ka, Divi\v{s}, Rudajevov\'{a}, and
  Sechovsk\'{y}}}]{YIr2Si2_2012}
\bibinfo{author}{\bibfnamefont{M.}~\bibnamefont{Vali\v{s}ka}},
  \bibinfo{author}{\bibfnamefont{J.}~\bibnamefont{Posp\'{\i}\v{s}il}},
  \bibinfo{author}{\bibfnamefont{J.}~\bibnamefont{Prokle\v{s}ka}},
  \bibinfo{author}{\bibfnamefont{M.}~\bibnamefont{Divi\v{s}}},
  \bibinfo{author}{\bibfnamefont{A.}~\bibnamefont{Rudajevov\'{a}}},
  \bibnamefont{and}
  \bibinfo{author}{\bibfnamefont{V.}~\bibnamefont{Sechovsk\'{y}}},
  \bibinfo{journal}{Journal of the Physical Society of Japan}
  \textbf{\bibinfo{volume}{81}}, \bibinfo{pages}{104715}
  (\bibinfo{year}{2012}), \eprint{https://doi.org/10.1143/JPSJ.81.104715},
  \urlprefix\url{https://doi.org/10.1143/JPSJ.81.104715}.

\bibitem[{\citenamefont{Yutaro et~al.}(2013)\citenamefont{Yutaro, Nobutaka,
  Akihiro, Hideki, Hirofumi, Masaki, Masahiko, and Yutaka}}]{Nagano2013}
\bibinfo{author}{\bibfnamefont{N.}~\bibnamefont{Yutaro}},
  \bibinfo{author}{\bibfnamefont{A.}~\bibnamefont{Nobutaka}},
  \bibinfo{author}{\bibfnamefont{M.}~\bibnamefont{Akihiro}},
  \bibinfo{author}{\bibfnamefont{Y.}~\bibnamefont{Hideki}},
  \bibinfo{author}{\bibfnamefont{W.}~\bibnamefont{Hirofumi}},
  \bibinfo{author}{\bibfnamefont{I.}~\bibnamefont{Masaki}},
  \bibinfo{author}{\bibfnamefont{I.}~\bibnamefont{Masahiko}}, \bibnamefont{and}
  \bibinfo{author}{\bibfnamefont{U.}~\bibnamefont{Yutaka}},
  \bibinfo{journal}{Journal of the Physical Society of Japan}
  \textbf{\bibinfo{volume}{82}}, \bibinfo{pages}{064715}
  (\bibinfo{year}{2013}), \eprint{https://doi.org/10.7566/JPSJ.82.064715},
  \urlprefix\url{https://doi.org/10.7566/JPSJ.82.064715}.

\bibitem[{\citenamefont{Gupta et~al.}(2017)\citenamefont{Gupta, Dhar,
  Thamizhavel, Rajeev, and Hossain}}]{Gupta_2017_La}
\bibinfo{author}{\bibfnamefont{R.}~\bibnamefont{Gupta}},
  \bibinfo{author}{\bibfnamefont{S.~K.} \bibnamefont{Dhar}},
  \bibinfo{author}{\bibfnamefont{A.}~\bibnamefont{Thamizhavel}},
  \bibinfo{author}{\bibfnamefont{K.~P.} \bibnamefont{Rajeev}},
  \bibnamefont{and} \bibinfo{author}{\bibfnamefont{Z.}~\bibnamefont{Hossain}},
  \bibinfo{journal}{Journal of Physics: Condensed Matter}
  \textbf{\bibinfo{volume}{29}}, \bibinfo{pages}{255601}
  (\bibinfo{year}{2017}),
  \urlprefix\url{https://dx.doi.org/10.1088/1361-648X/aa70a7}.

\bibitem[{\citenamefont{Falkowski et~al.}(2019)\citenamefont{Falkowski,
  Dole\ifmmode~\check{z}\else \v{z}\fi{}al, Andreev, Duverger-N\'edellec, and
  Havela}}]{Falkowski_2019_La}
\bibinfo{author}{\bibfnamefont{M.}~\bibnamefont{Falkowski}},
  \bibinfo{author}{\bibfnamefont{P.}~\bibnamefont{Dole\ifmmode~\check{z}\else
  \v{z}\fi{}al}}, \bibinfo{author}{\bibfnamefont{A.~V.} \bibnamefont{Andreev}},
  \bibinfo{author}{\bibfnamefont{E.}~\bibnamefont{Duverger-N\'edellec}},
  \bibnamefont{and} \bibinfo{author}{\bibfnamefont{L.}~\bibnamefont{Havela}},
  \bibinfo{journal}{Phys. Rev. B} \textbf{\bibinfo{volume}{100}},
  \bibinfo{pages}{064103} (\bibinfo{year}{2019}),
  \urlprefix\url{https://link.aps.org/doi/10.1103/PhysRevB.100.064103}.

\bibitem[{\citenamefont{Das et~al.}(2018)\citenamefont{Das, Gupta,
  Bhattacharyya, Biswas, Adroja, and Hossain}}]{La_muon_2018}
\bibinfo{author}{\bibfnamefont{D.}~\bibnamefont{Das}},
  \bibinfo{author}{\bibfnamefont{R.}~\bibnamefont{Gupta}},
  \bibinfo{author}{\bibfnamefont{A.}~\bibnamefont{Bhattacharyya}},
  \bibinfo{author}{\bibfnamefont{P.~K.} \bibnamefont{Biswas}},
  \bibinfo{author}{\bibfnamefont{D.~T.} \bibnamefont{Adroja}},
  \bibnamefont{and} \bibinfo{author}{\bibfnamefont{Z.}~\bibnamefont{Hossain}},
  \bibinfo{journal}{Phys. Rev. B} \textbf{\bibinfo{volume}{97}},
  \bibinfo{pages}{184509} (\bibinfo{year}{2018}),
  \urlprefix\url{https://link.aps.org/doi/10.1103/PhysRevB.97.184509}.

\bibitem[{\citenamefont{Shen et~al.}(2020)\citenamefont{Shen, Du, Li,
  Thamizhavel, Smidman, Nie, Luo, Le, Hossain, and Yuan}}]{La_pressure_2020}
\bibinfo{author}{\bibfnamefont{B.}~\bibnamefont{Shen}},
  \bibinfo{author}{\bibfnamefont{F.}~\bibnamefont{Du}},
  \bibinfo{author}{\bibfnamefont{R.}~\bibnamefont{Li}},
  \bibinfo{author}{\bibfnamefont{A.}~\bibnamefont{Thamizhavel}},
  \bibinfo{author}{\bibfnamefont{M.}~\bibnamefont{Smidman}},
  \bibinfo{author}{\bibfnamefont{Z.~Y.} \bibnamefont{Nie}},
  \bibinfo{author}{\bibfnamefont{S.~S.} \bibnamefont{Luo}},
  \bibinfo{author}{\bibfnamefont{T.}~\bibnamefont{Le}},
  \bibinfo{author}{\bibfnamefont{Z.}~\bibnamefont{Hossain}}, \bibnamefont{and}
  \bibinfo{author}{\bibfnamefont{H.~Q.} \bibnamefont{Yuan}},
  \bibinfo{journal}{Phys. Rev. B} \textbf{\bibinfo{volume}{101}},
  \bibinfo{pages}{144501} (\bibinfo{year}{2020}),
  \urlprefix\url{https://link.aps.org/doi/10.1103/PhysRevB.101.144501}.

\bibitem[{\citenamefont{Mukkattukavil et~al.}(2022)\citenamefont{Mukkattukavil,
  Hellsvik, Ghosh, Chatzigeorgiou, Nocerino, Wang, von Arx, Huang, Ekholm,
  Hossain et~al.}}]{La_RIXS_2022}
\bibinfo{author}{\bibfnamefont{D.~J.} \bibnamefont{Mukkattukavil}},
  \bibinfo{author}{\bibfnamefont{J.}~\bibnamefont{Hellsvik}},
  \bibinfo{author}{\bibfnamefont{A.}~\bibnamefont{Ghosh}},
  \bibinfo{author}{\bibfnamefont{E.}~\bibnamefont{Chatzigeorgiou}},
  \bibinfo{author}{\bibfnamefont{E.}~\bibnamefont{Nocerino}},
  \bibinfo{author}{\bibfnamefont{Q.}~\bibnamefont{Wang}},
  \bibinfo{author}{\bibfnamefont{K.}~\bibnamefont{von Arx}},
  \bibinfo{author}{\bibfnamefont{S.-W.} \bibnamefont{Huang}},
  \bibinfo{author}{\bibfnamefont{V.}~\bibnamefont{Ekholm}},
  \bibinfo{author}{\bibfnamefont{Z.}~\bibnamefont{Hossain}},
  \bibnamefont{et~al.}, \bibinfo{journal}{Journal of Physics: Condensed Matter}
  \textbf{\bibinfo{volume}{34}}, \bibinfo{pages}{324003}
  (\bibinfo{year}{2022}),
  \urlprefix\url{https://dx.doi.org/10.1088/1361-648X/ac7500}.

\bibitem[{\citenamefont{Nocerino et~al.}(2023)\citenamefont{Nocerino, Stuhr,
  {San Lorenzo}, Mazza, Mazzone, Hellsvik, Hasegawa, Asai, Masuda, Itoh
  et~al.}}]{La_syncrotron_NOCERINO2023100621}
\bibinfo{author}{\bibfnamefont{E.}~\bibnamefont{Nocerino}},
  \bibinfo{author}{\bibfnamefont{U.}~\bibnamefont{Stuhr}},
  \bibinfo{author}{\bibfnamefont{I.}~\bibnamefont{{San Lorenzo}}},
  \bibinfo{author}{\bibfnamefont{F.}~\bibnamefont{Mazza}},
  \bibinfo{author}{\bibfnamefont{D.}~\bibnamefont{Mazzone}},
  \bibinfo{author}{\bibfnamefont{J.}~\bibnamefont{Hellsvik}},
  \bibinfo{author}{\bibfnamefont{S.}~\bibnamefont{Hasegawa}},
  \bibinfo{author}{\bibfnamefont{S.}~\bibnamefont{Asai}},
  \bibinfo{author}{\bibfnamefont{T.}~\bibnamefont{Masuda}},
  \bibinfo{author}{\bibfnamefont{S.}~\bibnamefont{Itoh}}, \bibnamefont{et~al.},
  \bibinfo{journal}{Journal of Science: Advanced Materials and Devices}
  \textbf{\bibinfo{volume}{8}}, \bibinfo{pages}{100621} (\bibinfo{year}{2023}),
  ISSN \bibinfo{issn}{2468-2179},
  \urlprefix\url{https://www.sciencedirect.com/science/article/pii/S2468217923000904}.

\bibitem[{\citenamefont{Nie et~al.}(2021)\citenamefont{Nie, Yin, Thamizhavel,
  Wang, Shen, Che, Du, Hossain, Smidman, Lu et~al.}}]{Nie2021}
\bibinfo{author}{\bibfnamefont{Z.~Y.} \bibnamefont{Nie}},
  \bibinfo{author}{\bibfnamefont{L.~C.} \bibnamefont{Yin}},
  \bibinfo{author}{\bibfnamefont{A.}~\bibnamefont{Thamizhavel}},
  \bibinfo{author}{\bibfnamefont{A.}~\bibnamefont{Wang}},
  \bibinfo{author}{\bibfnamefont{B.}~\bibnamefont{Shen}},
  \bibinfo{author}{\bibfnamefont{L.~Q.} \bibnamefont{Che}},
  \bibinfo{author}{\bibfnamefont{F.}~\bibnamefont{Du}},
  \bibinfo{author}{\bibfnamefont{Z.}~\bibnamefont{Hossain}},
  \bibinfo{author}{\bibfnamefont{M.}~\bibnamefont{Smidman}},
  \bibinfo{author}{\bibfnamefont{X.}~\bibnamefont{Lu}}, \bibnamefont{et~al.},
  \bibinfo{journal}{Phys. Rev. B} \textbf{\bibinfo{volume}{103}},
  \bibinfo{pages}{014515} (\bibinfo{year}{2021}),
  \urlprefix\url{https://link.aps.org/doi/10.1103/PhysRevB.103.014515}.

\bibitem[{\citenamefont{Kudo et~al.}(2010)\citenamefont{Kudo, Nishikubo, and
  Nohara}}]{SrPt2As2_2010}
\bibinfo{author}{\bibfnamefont{K.}~\bibnamefont{Kudo}},
  \bibinfo{author}{\bibfnamefont{Y.}~\bibnamefont{Nishikubo}},
  \bibnamefont{and} \bibinfo{author}{\bibfnamefont{M.}~\bibnamefont{Nohara}},
  \bibinfo{journal}{Journal of the Physical Society of Japan}
  \textbf{\bibinfo{volume}{79}}, \bibinfo{pages}{123710}
  (\bibinfo{year}{2010}), \eprint{https://doi.org/10.1143/JPSJ.79.123710},
  \urlprefix\url{https://doi.org/10.1143/JPSJ.79.123710}.

\bibitem[{\citenamefont{Guo et~al.}(2016)\citenamefont{Guo, Jiang, Smidman,
  Han, Malliakas, Shen, Wang, Chen, Lu, Kanatzidis et~al.}}]{Guo2016}
\bibinfo{author}{\bibfnamefont{C.~Y.} \bibnamefont{Guo}},
  \bibinfo{author}{\bibfnamefont{W.~B.} \bibnamefont{Jiang}},
  \bibinfo{author}{\bibfnamefont{M.}~\bibnamefont{Smidman}},
  \bibinfo{author}{\bibfnamefont{F.}~\bibnamefont{Han}},
  \bibinfo{author}{\bibfnamefont{C.~D.} \bibnamefont{Malliakas}},
  \bibinfo{author}{\bibfnamefont{B.}~\bibnamefont{Shen}},
  \bibinfo{author}{\bibfnamefont{Y.~F.} \bibnamefont{Wang}},
  \bibinfo{author}{\bibfnamefont{Y.}~\bibnamefont{Chen}},
  \bibinfo{author}{\bibfnamefont{X.}~\bibnamefont{Lu}},
  \bibinfo{author}{\bibfnamefont{M.~G.} \bibnamefont{Kanatzidis}},
  \bibnamefont{et~al.}, \bibinfo{journal}{Phys. Rev. B}
  \textbf{\bibinfo{volume}{94}}, \bibinfo{pages}{184506}
  (\bibinfo{year}{2016}),
  \urlprefix\url{https://link.aps.org/doi/10.1103/PhysRevB.94.184506}.

\bibitem[{\citenamefont{Pikul et~al.}(2017)\citenamefont{Pikul,
  Samsel–Czekała, Chajewski, Romanova, Hackemer, Gorzelniak, Wiśniewski,
  and Kaczorowski}}]{Pikul_2017}
\bibinfo{author}{\bibfnamefont{A.~P.} \bibnamefont{Pikul}},
  \bibinfo{author}{\bibfnamefont{M.}~\bibnamefont{Samsel–Czekała}},
  \bibinfo{author}{\bibfnamefont{G.}~\bibnamefont{Chajewski}},
  \bibinfo{author}{\bibfnamefont{T.}~\bibnamefont{Romanova}},
  \bibinfo{author}{\bibfnamefont{A.}~\bibnamefont{Hackemer}},
  \bibinfo{author}{\bibfnamefont{R.}~\bibnamefont{Gorzelniak}},
  \bibinfo{author}{\bibfnamefont{P.}~\bibnamefont{Wiśniewski}},
  \bibnamefont{and}
  \bibinfo{author}{\bibfnamefont{D.}~\bibnamefont{Kaczorowski}},
  \bibinfo{journal}{Journal of Physics: Condensed Matter}
  \textbf{\bibinfo{volume}{29}}, \bibinfo{pages}{195602}
  (\bibinfo{year}{2017}),
  \urlprefix\url{https://dx.doi.org/10.1088/1361-648X/aa6832}.

\bibitem[{\citenamefont{Bl{\"o}chl}(1994)}]{blochl1994projector}
\bibinfo{author}{\bibfnamefont{P.~E.} \bibnamefont{Bl{\"o}chl}},
  \bibinfo{journal}{Physical review B} \textbf{\bibinfo{volume}{50}},
  \bibinfo{pages}{17953} (\bibinfo{year}{1994}).

\bibitem[{\citenamefont{Kresse and Joubert}(1999)}]{kresse1999ultrasoft}
\bibinfo{author}{\bibfnamefont{G.}~\bibnamefont{Kresse}} \bibnamefont{and}
  \bibinfo{author}{\bibfnamefont{D.}~\bibnamefont{Joubert}},
  \bibinfo{journal}{Physical review b} \textbf{\bibinfo{volume}{59}},
  \bibinfo{pages}{1758} (\bibinfo{year}{1999}).

\bibitem[{\citenamefont{Hobbs et~al.}(2000)\citenamefont{Hobbs, Kresse, and
  Hafner}}]{hobbs2000fully}
\bibinfo{author}{\bibfnamefont{D.}~\bibnamefont{Hobbs}},
  \bibinfo{author}{\bibfnamefont{G.}~\bibnamefont{Kresse}}, \bibnamefont{and}
  \bibinfo{author}{\bibfnamefont{J.}~\bibnamefont{Hafner}},
  \bibinfo{journal}{Physical Review B} \textbf{\bibinfo{volume}{62}},
  \bibinfo{pages}{11556} (\bibinfo{year}{2000}).

\bibitem[{\citenamefont{Perdew et~al.}(2008)\citenamefont{Perdew, Ruzsinszky,
  Csonka, Vydrov, Scuseria, Constantin, Zhou, and Burke}}]{perdew2008restoring}
\bibinfo{author}{\bibfnamefont{J.~P.} \bibnamefont{Perdew}},
  \bibinfo{author}{\bibfnamefont{A.}~\bibnamefont{Ruzsinszky}},
  \bibinfo{author}{\bibfnamefont{G.~I.} \bibnamefont{Csonka}},
  \bibinfo{author}{\bibfnamefont{O.~A.} \bibnamefont{Vydrov}},
  \bibinfo{author}{\bibfnamefont{G.~E.} \bibnamefont{Scuseria}},
  \bibinfo{author}{\bibfnamefont{L.~A.} \bibnamefont{Constantin}},
  \bibinfo{author}{\bibfnamefont{X.}~\bibnamefont{Zhou}}, \bibnamefont{and}
  \bibinfo{author}{\bibfnamefont{K.}~\bibnamefont{Burke}},
  \bibinfo{journal}{Physical review letters} \textbf{\bibinfo{volume}{100}},
  \bibinfo{pages}{136406} (\bibinfo{year}{2008}).

\bibitem[{\citenamefont{Togo et~al.}(2020)\citenamefont{Togo, Inoue, and
  Tanaka}}]{togo2020phonon}
\bibinfo{author}{\bibfnamefont{A.}~\bibnamefont{Togo}},
  \bibinfo{author}{\bibfnamefont{Y.}~\bibnamefont{Inoue}}, \bibnamefont{and}
  \bibinfo{author}{\bibfnamefont{I.}~\bibnamefont{Tanaka}},
  \bibinfo{journal}{Physical Review B} \textbf{\bibinfo{volume}{102}},
  \bibinfo{pages}{024106} (\bibinfo{year}{2020}).

\bibitem[{\citenamefont{Hiebl and Rogl}(1985)}]{HIEBL1985}
\bibinfo{author}{\bibfnamefont{K.}~\bibnamefont{Hiebl}} \bibnamefont{and}
  \bibinfo{author}{\bibfnamefont{P.}~\bibnamefont{Rogl}},
  \bibinfo{journal}{Journal of Magnetism and Magnetic Materials}
  \textbf{\bibinfo{volume}{50}}, \bibinfo{pages}{39} (\bibinfo{year}{1985}),
  ISSN \bibinfo{issn}{0304-8853},
  \urlprefix\url{https://www.sciencedirect.com/science/article/pii/0304885385900848}.

\bibitem[{\citenamefont{Samsel–Czekała
  et~al.}(2018)\citenamefont{Samsel–Czekała, Chajewski, Wiśniewski,
  Romanova, Hackemer, Gorzelniak, Pikul, and Kaczorowski}}]{LuT2Si2_2018}
\bibinfo{author}{\bibfnamefont{M.}~\bibnamefont{Samsel–Czekała}},
  \bibinfo{author}{\bibfnamefont{G.}~\bibnamefont{Chajewski}},
  \bibinfo{author}{\bibfnamefont{P.}~\bibnamefont{Wiśniewski}},
  \bibinfo{author}{\bibfnamefont{T.}~\bibnamefont{Romanova}},
  \bibinfo{author}{\bibfnamefont{A.}~\bibnamefont{Hackemer}},
  \bibinfo{author}{\bibfnamefont{R.}~\bibnamefont{Gorzelniak}},
  \bibinfo{author}{\bibfnamefont{A.}~\bibnamefont{Pikul}}, \bibnamefont{and}
  \bibinfo{author}{\bibfnamefont{D.}~\bibnamefont{Kaczorowski}},
  \bibinfo{journal}{Physica B: Condensed Matter}
  \textbf{\bibinfo{volume}{536}}, \bibinfo{pages}{816} (\bibinfo{year}{2018}),
  ISSN \bibinfo{issn}{0921-4526},
  \urlprefix\url{https://www.sciencedirect.com/science/article/pii/S0921452617307305}.

\bibitem[{\citenamefont{Fernandes et~al.}(2022)\citenamefont{Fernandes, Coldea,
  Ding, Fisher, Hirschfeld, and Kotliar}}]{Rafael2022}
\bibinfo{author}{\bibfnamefont{R.~M.} \bibnamefont{Fernandes}},
  \bibinfo{author}{\bibfnamefont{A.~I.} \bibnamefont{Coldea}},
  \bibinfo{author}{\bibfnamefont{H.}~\bibnamefont{Ding}},
  \bibinfo{author}{\bibfnamefont{I.~R.} \bibnamefont{Fisher}},
  \bibinfo{author}{\bibfnamefont{P.~J.} \bibnamefont{Hirschfeld}},
  \bibnamefont{and} \bibinfo{author}{\bibfnamefont{G.}~\bibnamefont{Kotliar}},
  \bibinfo{journal}{Nature (London)} \textbf{\bibinfo{volume}{601}},
  \bibinfo{pages}{35} (\bibinfo{year}{2022}),
  \urlprefix\url{https://www.nature.com/articles/s41586-021-04073-2#citeas}.

\bibitem[{\citenamefont{Hwang and Das~Sarma}(2019)}]{Hwang_2019}
\bibinfo{author}{\bibfnamefont{E.~H.} \bibnamefont{Hwang}} \bibnamefont{and}
  \bibinfo{author}{\bibfnamefont{S.}~\bibnamefont{Das~Sarma}},
  \bibinfo{journal}{Phys. Rev. B} \textbf{\bibinfo{volume}{99}},
  \bibinfo{pages}{085105} (\bibinfo{year}{2019}),
  \urlprefix\url{https://link.aps.org/doi/10.1103/PhysRevB.99.085105}.

\bibitem[{\citenamefont{Ziman}(2001)}]{ziman2001electrons}
\bibinfo{author}{\bibfnamefont{J.}~\bibnamefont{Ziman}},
  \emph{\bibinfo{title}{Electrons and Phonons: The Theory of Transport
  Phenomena in Solids}}, International series of monographs on physics
  (\bibinfo{publisher}{OUP Oxford}, \bibinfo{year}{2001}), ISBN
  \bibinfo{isbn}{9780198507796},
  \urlprefix\url{https://books.google.com.br/books?id=UtEy63pjngsC}.

\bibitem[{\citenamefont{Bruin et~al.}(2013)\citenamefont{Bruin, Sakai, Perry,
  and Mackenzie}}]{Bruin_2013}
\bibinfo{author}{\bibfnamefont{J.~A.~N.} \bibnamefont{Bruin}},
  \bibinfo{author}{\bibfnamefont{H.}~\bibnamefont{Sakai}},
  \bibinfo{author}{\bibfnamefont{R.~S.} \bibnamefont{Perry}}, \bibnamefont{and}
  \bibinfo{author}{\bibfnamefont{A.~P.} \bibnamefont{Mackenzie}},
  \bibinfo{journal}{Science} \textbf{\bibinfo{volume}{339}},
  \bibinfo{pages}{804} (\bibinfo{year}{2013}),
  \urlprefix\url{https://www.science.org/doi/abs/10.1126/science.1227612}.

\bibitem[{\citenamefont{Yang et~al.}(2022)\citenamefont{Yang, Yang, Su, Fang,
  Yang, Chen, Wang, Du, Wu, and Fang}}]{Yang_2022}
\bibinfo{author}{\bibfnamefont{Z.}~\bibnamefont{Yang}},
  \bibinfo{author}{\bibfnamefont{Z.}~\bibnamefont{Yang}},
  \bibinfo{author}{\bibfnamefont{Q.}~\bibnamefont{Su}},
  \bibinfo{author}{\bibfnamefont{E.}~\bibnamefont{Fang}},
  \bibinfo{author}{\bibfnamefont{J.}~\bibnamefont{Yang}},
  \bibinfo{author}{\bibfnamefont{B.}~\bibnamefont{Chen}},
  \bibinfo{author}{\bibfnamefont{H.}~\bibnamefont{Wang}},
  \bibinfo{author}{\bibfnamefont{J.}~\bibnamefont{Du}},
  \bibinfo{author}{\bibfnamefont{C.}~\bibnamefont{Wu}}, \bibnamefont{and}
  \bibinfo{author}{\bibfnamefont{M.}~\bibnamefont{Fang}},
  \bibinfo{journal}{Phys. Rev. B} \textbf{\bibinfo{volume}{106}},
  \bibinfo{pages}{224501} (\bibinfo{year}{2022}),
  \urlprefix\url{https://link.aps.org/doi/10.1103/PhysRevB.106.224501}.

\bibitem[{\citenamefont{Takayama et~al.}(2012)\citenamefont{Takayama, Kuwano,
  Hirai, Katsura, Yamamoto, and Takagi}}]{Takayama_2012}
\bibinfo{author}{\bibfnamefont{T.}~\bibnamefont{Takayama}},
  \bibinfo{author}{\bibfnamefont{K.}~\bibnamefont{Kuwano}},
  \bibinfo{author}{\bibfnamefont{D.}~\bibnamefont{Hirai}},
  \bibinfo{author}{\bibfnamefont{Y.}~\bibnamefont{Katsura}},
  \bibinfo{author}{\bibfnamefont{A.}~\bibnamefont{Yamamoto}}, \bibnamefont{and}
  \bibinfo{author}{\bibfnamefont{H.}~\bibnamefont{Takagi}},
  \bibinfo{journal}{Phys. Rev. Lett.} \textbf{\bibinfo{volume}{108}},
  \bibinfo{pages}{237001} (\bibinfo{year}{2012}),
  \urlprefix\url{https://link.aps.org/doi/10.1103/PhysRevLett.108.237001}.

\bibitem[{\citenamefont{Sundar et~al.}(2019)\citenamefont{Sundar, Salem-Sugui,
  Chattopadhyay, Roy, Sharath~Chandra, Cohen, and Ghivelder}}]{Sundar_2019}
\bibinfo{author}{\bibfnamefont{S.}~\bibnamefont{Sundar}},
  \bibinfo{author}{\bibfnamefont{S.}~\bibnamefont{Salem-Sugui}},
  \bibinfo{author}{\bibfnamefont{M.~K.} \bibnamefont{Chattopadhyay}},
  \bibinfo{author}{\bibfnamefont{S.~B.} \bibnamefont{Roy}},
  \bibinfo{author}{\bibfnamefont{L.~S.} \bibnamefont{Sharath~Chandra}},
  \bibinfo{author}{\bibfnamefont{L.~F.} \bibnamefont{Cohen}}, \bibnamefont{and}
  \bibinfo{author}{\bibfnamefont{L.}~\bibnamefont{Ghivelder}},
  \bibinfo{journal}{Superconductor Science and Technology}
  \textbf{\bibinfo{volume}{32}}, \bibinfo{pages}{055003}
  (\bibinfo{year}{2019}),
  \urlprefix\url{https://doi.org/10.1088/1361-6668/ab06a5}.

\bibitem[{\citenamefont{Yu and Anderson}(1984)}]{Yu_1984}
\bibinfo{author}{\bibfnamefont{C.~C.} \bibnamefont{Yu}} \bibnamefont{and}
  \bibinfo{author}{\bibfnamefont{P.~W.} \bibnamefont{Anderson}},
  \bibinfo{journal}{Phys. Rev. B} \textbf{\bibinfo{volume}{29}},
  \bibinfo{pages}{6165} (\bibinfo{year}{1984}),
  \urlprefix\url{https://link.aps.org/doi/10.1103/PhysRevB.29.6165}.

\bibitem[{\citenamefont{Matsuura and Miyake}(1986)}]{Matsuura_1986}
\bibinfo{author}{\bibfnamefont{T.}~\bibnamefont{Matsuura}} \bibnamefont{and}
  \bibinfo{author}{\bibfnamefont{K.}~\bibnamefont{Miyake}},
  \bibinfo{journal}{Journal of the Physical Society of Japan}
  \textbf{\bibinfo{volume}{55}}, \bibinfo{pages}{610} (\bibinfo{year}{1986}),
  \eprint{https://doi.org/10.1143/JPSJ.55.610},
  \urlprefix\url{https://doi.org/10.1143/JPSJ.55.610}.

\bibitem[{\citenamefont{Jacko et~al.}(2009)\citenamefont{Jacko, Fj{\ae}restad,
  and Powell}}]{KWR_Jacko2009}
\bibinfo{author}{\bibfnamefont{A.~C.} \bibnamefont{Jacko}},
  \bibinfo{author}{\bibfnamefont{J.~O.} \bibnamefont{Fj{\ae}restad}},
  \bibnamefont{and} \bibinfo{author}{\bibfnamefont{B.~J.}
  \bibnamefont{Powell}}, \bibinfo{journal}{Nature Physics}
  \textbf{\bibinfo{volume}{5}}, \bibinfo{pages}{422} (\bibinfo{year}{2009}),
  ISSN \bibinfo{issn}{1745-2481},
  \urlprefix\url{https://doi.org/10.1038/nphys1249}.

\bibitem[{\citenamefont{Tari}(2003)}]{tari2003specific}
\bibinfo{author}{\bibfnamefont{A.}~\bibnamefont{Tari}},
  \emph{\bibinfo{title}{The Specific Heat of Matter at Low Temperatures}}
  (\bibinfo{publisher}{Imperial College Press}, \bibinfo{year}{2003}), ISBN
  \bibinfo{isbn}{9781860943140},
  \urlprefix\url{https://books.google.com.br/books?id=ymFQ0pRezKwC}.

\bibitem[{\citenamefont{Orlando et~al.}(1979)\citenamefont{Orlando, McNiff,
  Foner, and Beasley}}]{SC_params_PhysRevB.19.4545}
\bibinfo{author}{\bibfnamefont{T.~P.} \bibnamefont{Orlando}},
  \bibinfo{author}{\bibfnamefont{E.~J.} \bibnamefont{McNiff}},
  \bibinfo{author}{\bibfnamefont{S.}~\bibnamefont{Foner}}, \bibnamefont{and}
  \bibinfo{author}{\bibfnamefont{M.~R.} \bibnamefont{Beasley}},
  \bibinfo{journal}{Phys. Rev. B} \textbf{\bibinfo{volume}{19}},
  \bibinfo{pages}{4545} (\bibinfo{year}{1979}),
  \urlprefix\url{https://link.aps.org/doi/10.1103/PhysRevB.19.4545}.

\bibitem[{\citenamefont{de~Gennes}(1966)}]{de1966superconductivity}
\bibinfo{author}{\bibfnamefont{P.}~\bibnamefont{de~Gennes}},
  \emph{\bibinfo{title}{Superconductivity of Metals and Alloys}}, Frontiers in
  physics (\bibinfo{publisher}{W.A. Benjamin}, \bibinfo{year}{1966}),
  \urlprefix\url{https://books.google.com.br/books?id=KtJEAAAAIAAJ}.

\bibitem[{\citenamefont{McMillan}(1968)}]{Mcmillan_1968}
\bibinfo{author}{\bibfnamefont{W.~L.} \bibnamefont{McMillan}},
  \bibinfo{journal}{Phys. Rev.} \textbf{\bibinfo{volume}{167}},
  \bibinfo{pages}{331} (\bibinfo{year}{1968}),
  \urlprefix\url{https://link.aps.org/doi/10.1103/PhysRev.167.331}.

\bibitem[{\citenamefont{Djied et~al.}(2014)\citenamefont{Djied, Khachai,
  Seddik, Khenata, Bouhemadou, Guechi, Murtaza, Bin-Omran, Alahmed, and
  Ameri}}]{djied2014structural}
\bibinfo{author}{\bibfnamefont{A.}~\bibnamefont{Djied}},
  \bibinfo{author}{\bibfnamefont{H.}~\bibnamefont{Khachai}},
  \bibinfo{author}{\bibfnamefont{T.}~\bibnamefont{Seddik}},
  \bibinfo{author}{\bibfnamefont{R.}~\bibnamefont{Khenata}},
  \bibinfo{author}{\bibfnamefont{A.}~\bibnamefont{Bouhemadou}},
  \bibinfo{author}{\bibfnamefont{N.}~\bibnamefont{Guechi}},
  \bibinfo{author}{\bibfnamefont{G.}~\bibnamefont{Murtaza}},
  \bibinfo{author}{\bibfnamefont{S.}~\bibnamefont{Bin-Omran}},
  \bibinfo{author}{\bibfnamefont{Z.}~\bibnamefont{Alahmed}}, \bibnamefont{and}
  \bibinfo{author}{\bibfnamefont{M.}~\bibnamefont{Ameri}},
  \bibinfo{journal}{Computational materials science}
  \textbf{\bibinfo{volume}{84}}, \bibinfo{pages}{396} (\bibinfo{year}{2014}).

\bibitem[{\citenamefont{Wu et~al.}(2007)\citenamefont{Wu, Zhao, Xiang, Hao,
  Liu, and Meng}}]{wu2007crystal}
\bibinfo{author}{\bibfnamefont{Z.-j.} \bibnamefont{Wu}},
  \bibinfo{author}{\bibfnamefont{E.-j.} \bibnamefont{Zhao}},
  \bibinfo{author}{\bibfnamefont{H.-p.} \bibnamefont{Xiang}},
  \bibinfo{author}{\bibfnamefont{X.-f.} \bibnamefont{Hao}},
  \bibinfo{author}{\bibfnamefont{X.-j.} \bibnamefont{Liu}}, \bibnamefont{and}
  \bibinfo{author}{\bibfnamefont{J.}~\bibnamefont{Meng}},
  \bibinfo{journal}{Physical Review B—Condensed Matter and Materials Physics}
  \textbf{\bibinfo{volume}{76}}, \bibinfo{pages}{054115}
  (\bibinfo{year}{2007}).

\bibitem[{\citenamefont{Becke and Edgecombe}(1990)}]{becke1990simple}
\bibinfo{author}{\bibfnamefont{A.~D.} \bibnamefont{Becke}} \bibnamefont{and}
  \bibinfo{author}{\bibfnamefont{K.~E.} \bibnamefont{Edgecombe}},
  \bibinfo{journal}{The Journal of chemical physics}
  \textbf{\bibinfo{volume}{92}}, \bibinfo{pages}{5397} (\bibinfo{year}{1990}).

\bibitem[{\citenamefont{Savin et~al.}(1997)\citenamefont{Savin, Nesper,
  Wengert, and F{\"a}ssler}}]{savin1997elf}
\bibinfo{author}{\bibfnamefont{A.}~\bibnamefont{Savin}},
  \bibinfo{author}{\bibfnamefont{R.}~\bibnamefont{Nesper}},
  \bibinfo{author}{\bibfnamefont{S.}~\bibnamefont{Wengert}}, \bibnamefont{and}
  \bibinfo{author}{\bibfnamefont{T.~F.} \bibnamefont{F{\"a}ssler}},
  \bibinfo{journal}{Angewandte Chemie International Edition in English}
  \textbf{\bibinfo{volume}{36}}, \bibinfo{pages}{1808} (\bibinfo{year}{1997}).

\bibitem[{\citenamefont{S. et~al.}(2015)\citenamefont{S., K., and
  B.}}]{kim_2015}
\bibinfo{author}{\bibfnamefont{K.}~\bibnamefont{S.}},
  \bibinfo{author}{\bibfnamefont{K.}~\bibnamefont{K.}}, \bibnamefont{and}
  \bibinfo{author}{\bibfnamefont{M.}~\bibnamefont{B.}}, \bibinfo{journal}{Sci
  Rep} \textbf{\bibinfo{volume}{5}} (\bibinfo{year}{2015}),
  \urlprefix\url{https://doi.org/10.1038/srep15052}.

\bibitem[{\citenamefont{Gupta et~al.}(2022)\citenamefont{Gupta, Thamizhavel,
  Rajeev, and Hossain}}]{Gupta_2022}
\bibinfo{author}{\bibfnamefont{R.}~\bibnamefont{Gupta}},
  \bibinfo{author}{\bibfnamefont{A.}~\bibnamefont{Thamizhavel}},
  \bibinfo{author}{\bibfnamefont{K.~P.} \bibnamefont{Rajeev}},
  \bibnamefont{and} \bibinfo{author}{\bibfnamefont{Z.}~\bibnamefont{Hossain}},
  \bibinfo{journal}{Superconductor Science and Technology}
  \textbf{\bibinfo{volume}{35}}, \bibinfo{pages}{084006}
  (\bibinfo{year}{2022}),
  \urlprefix\url{https://doi.org/10.1088/1361-6668/ac7755}.

\bibitem[{\citenamefont{Junod et~al.}(1979)\citenamefont{Junod, Bichsel, and
  Muller}}]{junod1979modification}
\bibinfo{author}{\bibfnamefont{A.}~\bibnamefont{Junod}},
  \bibinfo{author}{\bibfnamefont{D.}~\bibnamefont{Bichsel}}, \bibnamefont{and}
  \bibinfo{author}{\bibfnamefont{J.}~\bibnamefont{Muller}},
  \bibinfo{journal}{Helvetica Physica Acta} \textbf{\bibinfo{volume}{52}},
  \bibinfo{pages}{580} (\bibinfo{year}{1979}).

\bibitem[{\citenamefont{Zhu et~al.}(2015)\citenamefont{Zhu, Cao, Zhang,
  Plummer, and Guo}}]{class_cdw}
\bibinfo{author}{\bibfnamefont{X.}~\bibnamefont{Zhu}},
  \bibinfo{author}{\bibfnamefont{Y.}~\bibnamefont{Cao}},
  \bibinfo{author}{\bibfnamefont{J.}~\bibnamefont{Zhang}},
  \bibinfo{author}{\bibfnamefont{E.~W.} \bibnamefont{Plummer}},
  \bibnamefont{and} \bibinfo{author}{\bibfnamefont{J.}~\bibnamefont{Guo}},
  \bibinfo{journal}{Proceedings of the National Academy of Sciences}
  \textbf{\bibinfo{volume}{112}}, \bibinfo{pages}{2367} (\bibinfo{year}{2015}),
  \eprint{https://www.pnas.org/doi/pdf/10.1073/pnas.1424791112},
  \urlprefix\url{https://www.pnas.org/doi/abs/10.1073/pnas.1424791112}.

\bibitem[{\citenamefont{Thorne}(1996)}]{CDW_thorne_1996}
\bibinfo{author}{\bibfnamefont{R.~E.} \bibnamefont{Thorne}},
  \bibinfo{journal}{Physics Today} \textbf{\bibinfo{volume}{49}},
  \bibinfo{pages}{42} (\bibinfo{year}{1996}), ISSN \bibinfo{issn}{0031-9228},
  \eprint{https://pubs.aip.org/physicstoday/article-pdf/49/5/42/8309683/42\_1\_online.pdf},
  \urlprefix\url{https://doi.org/10.1063/1.881498}.

\bibitem[{\citenamefont{Petkov et~al.}(2023)\citenamefont{Petkov, Baumbach,
  Milinda~Abeykoon, and Mydosh}}]{UPt2Si2_2023}
\bibinfo{author}{\bibfnamefont{V.}~\bibnamefont{Petkov}},
  \bibinfo{author}{\bibfnamefont{R.}~\bibnamefont{Baumbach}},
  \bibinfo{author}{\bibfnamefont{A.~M.} \bibnamefont{Milinda~Abeykoon}},
  \bibnamefont{and} \bibinfo{author}{\bibfnamefont{J.~A.}
  \bibnamefont{Mydosh}}, \bibinfo{journal}{Phys. Rev. B}
  \textbf{\bibinfo{volume}{107}}, \bibinfo{pages}{245101}
  (\bibinfo{year}{2023}),
  \urlprefix\url{https://link.aps.org/doi/10.1103/PhysRevB.107.245101}.

\bibitem[{\citenamefont{Falkowski et~al.}(2020)\citenamefont{Falkowski,
  Dole\ifmmode~\check{z}\else \v{z}\fi{}al, Duverger-N\'edellec, Chamoreau,
  Fort\'e, Andreev, and Havela}}]{NdPt2Si2_2020}
\bibinfo{author}{\bibfnamefont{M.}~\bibnamefont{Falkowski}},
  \bibinfo{author}{\bibfnamefont{P.}~\bibnamefont{Dole\ifmmode~\check{z}\else
  \v{z}\fi{}al}},
  \bibinfo{author}{\bibfnamefont{E.}~\bibnamefont{Duverger-N\'edellec}},
  \bibinfo{author}{\bibfnamefont{L.-M.} \bibnamefont{Chamoreau}},
  \bibinfo{author}{\bibfnamefont{J.}~\bibnamefont{Fort\'e}},
  \bibinfo{author}{\bibfnamefont{A.~V.} \bibnamefont{Andreev}},
  \bibnamefont{and} \bibinfo{author}{\bibfnamefont{L.}~\bibnamefont{Havela}},
  \bibinfo{journal}{Phys. Rev. B} \textbf{\bibinfo{volume}{101}},
  \bibinfo{pages}{174110} (\bibinfo{year}{2020}),
  \urlprefix\url{https://link.aps.org/doi/10.1103/PhysRevB.101.174110}.

\bibitem[{\citenamefont{Kumar et~al.}(2010)\citenamefont{Kumar, Anand, Geibel,
  Nicklas, and Hossain}}]{Kumar2010}
\bibinfo{author}{\bibfnamefont{M.}~\bibnamefont{Kumar}},
  \bibinfo{author}{\bibfnamefont{V.~K.} \bibnamefont{Anand}},
  \bibinfo{author}{\bibfnamefont{C.}~\bibnamefont{Geibel}},
  \bibinfo{author}{\bibfnamefont{M.}~\bibnamefont{Nicklas}}, \bibnamefont{and}
  \bibinfo{author}{\bibfnamefont{Z.}~\bibnamefont{Hossain}},
  \bibinfo{journal}{Phys. Rev. B} \textbf{\bibinfo{volume}{81}},
  \bibinfo{pages}{125107} (\bibinfo{year}{2010}),
  \urlprefix\url{https://link.aps.org/doi/10.1103/PhysRevB.81.125107}.

\end{thebibliography}




\end{document}